\documentclass[a4paper,11pt]{article}

\pdfoutput=1
\usepackage{jheppub}
\usepackage[T1]{fontenc}
\usepackage{graphicx,amstext,amsmath,amssymb,epsfig,color,ulem,color,bbold}
\newcommand{\eqn}[2] {\begin{equation}\label{#1}{#2}\end{equation}}
\newcommand{\eno}[1] {Eq.~\eqref{#1}}
\newcommand{\gsim}{\lower.7ex\hbox{$\;\stackrel{\textstyle>}{\sim}\;$}}
\newcommand{\lsim}{\lower.7ex\hbox{$\;\stackrel{\textstyle<}{\sim}\;$}}

\newcommand{\hs}{\hspace{0.12mm}}
\newcommand{\cA}{{\cal A}}

\def\N{{\mathcal{N}}}
\def\lGB{{\lambda_{\rm GB}}}

\usepackage{mathtools}
\usepackage{hyperref} 
\usepackage{soul}
\hypersetup{
    unicode=true,          
    bookmarksnumbered=true,
    colorlinks=true,       
    linkcolor=red,          
    citecolor=[rgb]{0,.7,0}, 
    filecolor=magenta,
    urlcolor=cyan,
    menucolor=blue,
    baseurl=http://www.arxiv.org/abs/,
    linktocpage=true,
    hyperfootnotes=true
}

\title{\boldmath Holographic Isotropisation in Gauss-Bonnet Gravity}

\author{Tom\'{a}s Andrade,}
\author{Jorge Casalderrey-Solana}
\author{and Andrej Ficnar}
\affiliation{Rudolf Peierls Centre for Theoretical Physics, University of Oxford,\\1 Keble Road, Oxford OX1 3NP, United Kingdom}

\emailAdd{tomas.andrade@physics.ox.ac.uk}
\emailAdd{jorge.casalderreysolana@physics.ox.ac.uk}
\emailAdd{andrej.ficnar@physics.ox.ac.uk}

\abstract{We study holographic isotropisation of homogeneous, strongly coupled, non-Abelian plasmas in Gauss-Bonnet gravity with a negative cosmological constant. We focus on  small values of the Gauss-Bonnet coupling parameter $\lambda_{GB}$ and linearise the equations of motion around a time-dependent background solution with $\lambda_{GB}=0$. We numerically solve the linearised equations and show that the entire time evolution of the pressure anisotropy can be well approximated by the linear in $\lambda_{GB}$ corrections to the quasinormal mode expansion, even in the cases of high anisotropy. We finally show that, quite generally, the time evolution of the pressure anisotropy with the Gauss-Bonnet term is approximately {\it shifted} with respect to the evolution without it, with the sign of the shift being directly related to the sign of the $\lambda_{GB}$ parameter. This suggests that finite coupling corrections generically {\it increase} the isotropisation time of strongly coupled plasmas. }

\begin{document} 
\maketitle
\flushbottom


\section{Introduction}
\label{Intro}

The gauge/string duality is a powerful theoretical laboratory in which to study the dynamics of strongly coupled, non-Abelian gauge theories. In the limit of large number of colours ($N_c$) and large 't Hooft coupling ($\lambda$), the duality maps complicated, fully quantum computations on the gauge theory side into simple and tractable classical gravity problems. This access to the strongly coupled sector of a large class of non-Abelian theories has led to many insights into the physics of strongly coupled matter at very different energy scales, from deconfined QCD matter to the behaviour of non-Fermi liquids, superconductors and cold atom systems (see \cite{Hartnoll:2008kx,McGreevy:2009xe,CasalderreySolana:2011us,Iqbal:2011ae} for recent reviews).

One of the areas of physics in which the gauge/string duality is most powerful is the analysis of far-from-equilibrium dynamics. Gravity computations have been employed to study, among many other subjects, the reaction of non-Abelian gauge theories to quenches \cite{Balasubramanian:2010ce,Buchel:2013lla,Buchel:2013gba,Bellantuono:2015hxa,Bellantuono:2016tkh}, the formation of turbulence \cite{Eling:2010vr,Adams:2012pj,Adams:2013vsa,Yang:2014tla}, the approach towards hydrodynamic behaviour of large disturbances of non-Abelian theories \cite{Janik:2005zt,Chesler:2008hg,Heller:2011ju,HoloLinPRL,HoloLinJHEP,Fuini:2015hba}, as well as the characterisation of the debris of the collision of energetic projectiles \cite{Chesler:2010bi,Casalderrey-Solana:2013aba,Chesler:2015wra,Casalderrey-Solana:2016xfq,CYReview}.  Nevertheless, most of the analyses performed up to date focus on the infinitely strongly coupled limit and very little is known about finite coupling effects to those dynamics. In this work we present the first step towards understanding finite coupling corrections to off-equilibrium dynamics. 

According to the holographic dictionary, finite coupling corrections in the gauge theory correspond to high curvature corrections on the gravity side. For the particular case of $\mathcal{N}=4$ supersymmetric Yang-Mills (SYM) theory, the complete set of high curvature, $R^4$, terms responsible for leading order corrections in the inverse 't Hooft coupling are known \cite{Gubser:1998nz}. These have been used in the past to determine the corrections to equilibrium properties of $\N=4$ SYM, such as the free energy \cite{Gubser:1998nz}, to the transport properties \cite{Buchel:2004di,Hassanain:2011fn,Grozdanov:2014kva}, as well as the equilibrium photon emission rate of the plasma \cite{Hassanain:2011ce}.  Corrections to near-equilibrium dynamics have been also addressed \cite{Stricker:2013lma,Steineder:2013ana,AndreiGB}, by computing the relaxation rates of small deviations from equilibrium, which on the gravity side are controlled by characteristic relaxation modes of black branes, known as  quasinormal modes (QNM) \cite{Berti:2009kk}.
Higher curvature corrections to thermalisation dynamics within a specific holographic construction in which AdS-Vadya black holes model the thermalisation after a sudden injection of energy,  have also been studied \cite{Steineder:2012si,Stricker:2013lma,Steineder:2013ana,Zeng:2013mca,Dey:2015poa}. Quadratic curvature effects in the off-equilibrium dynamics of homogenous matter in an expanding universe have been also recently considered in \cite{Camilo:2016kxq}.
 For a compilation of finite coupling corrections to infinitely strongly coupled $\N=4$ SYM see \cite{Waeber:2015oka}. 

In generic holographic constructions, one expects the leading higher curvature correction to be quadratic \cite{Boulware:1985wk}. A particular model of this kind is Gauss-Bonnet gravity, which is based on a particular set of $R^2$ corrections to the Einstein-Hilbert action with negative cosmological constant controlled by a single parameter $\lGB$ \cite{Nojiri:1999mh, Blau:1999vz, Cai-GBSoln, Kats:2007mq,Buchel:2008vz,Myers-GB, Buchel:2009tt}. While on the field theory side the holographic dual of Gauss-Bonnet gravity is unknown, this setup is appealing since the equations of motion remain of second order, so in principle we expect the theory to be free of the pathologies induced by generic higher derivative terms.\footnote{The inclusion of higher curvature corrections in the holographic context 
has been called into question by \cite{Camanho:2014apa}, which found that certain causality pathologies arise in the graviton three point functions of such theories unless one includes the full tower of corrections coming from a stringy model; see however \cite{Papallo:2015rna}.} In addition, unlike the higher curvature corrections of $\N=4$, the black hole solution of the Gauss-Bonnet theory with negative cosmological constant can be found analytically \cite{Cai-GBSoln}, so some calculations are amenable to treatment beyond perturbation theory in $\lambda_{GB}$ in a convenient way. Recently, the analysis of the relaxation of small out-of-equilibrium perturbations for arbitrary values of $\lGB$ has been performed in \cite{AndreiGB}. This revealed a very interesting behaviour of the relaxation for large values of the $\lGB$ parameter, which qualitatively resembled the expected behaviour of $\mathcal{N}=4$ SYM plasma in the small 't Hooft coupling limit. Furthermore, for  small negative values of $\lGB$, the authors of \cite{AndreiGB} found that the structure of small corrections to the QNM spectrum is qualitatively similar to that of infinitesimal $1 / \sqrt\lambda$ corrections to the relaxation dynamics in strongly coupled $\mathcal{N}=4$ SYM computed in \cite{Steineder:2012si,Stricker:2013lma,AndreiGB}. These facts indicate that Gauss-Bonnet gravity  provides a good testing ground for understanding the effects of finite coupling corrections on dynamical processes in strongly coupled $\N=4$ SYM.

In this paper, we study the problem of isotropisation in Gauss-Bonnet gravity. At an initial time, we prepare a homogeneous, anisotropic off-equilibrium state of the field theory dual to Gauss-Bonnet gravity, and study the process by which the system becomes isotropic. The initial disturbances are large, meaning that the difference in pressure between the anisotropic direction and the transverse directions is large in units of the energy density, which is constant as a consequence of the homogeneity of the state. On the gravity side, this setup corresponds to specifying the dual metric on the initial time-slice and using Einstein's equation to evolve the different metric fields. In the limit of small $\lGB$, we will consider the effect of higher curvature corrections as a small (linear) perturbation on top the non-linear evolution of the metric at $\lGB=0$. In this way, we extract the leading order correction in $\lGB$ to the isotropisation dynamics of the dual field theory.

The problem of non-linear isotropisation in $\N=4$ SYM has been studied in the past \cite{Chesler:2008hg, HoloLinPRL,HoloLinJHEP,Fuini:2015hba,CYReview}. Quite remarkably, those analyses have shown that, in spite of the intrinsically non-linear setup, the full out-of-equilibirum dynamics of this strongly coupled theory can be described as a linear superposition of the relaxation modes of the static black brane dual to the thermal system for a large class of initial conditions.  This observation provides a tremendous simplification, since it renders the complicated dynamics of isotropisation into a linear problem. In this paper we will find that in the presence of higher curvature corrections this simplification is preserved for moderately large ($\mathcal{O}(1)$) anisotropies, although not for arbitrarily large ones. As we will show, for the initial configurations described above, the evolution of the perturbed plasma can be approximated as a linear combination of the QNM of the black brane solution in Gauss-Bonnet gravity.

We also investigate the effect of the Gauss-Bonnet term on isotropisation time, defined as the time at which the ratio of the pressure anisotropy to the average pressure relaxes beyond a given threshold criterion. As observed in previous studies \cite{HoloLinPRL,HoloLinJHEP,Fuini:2015hba,CYReview}, the isotropisation time is not unique, but depends on the initial configuration. Similarly, the corrections induced by the Gauss-Bonnet term depend on the initial perturbation. Nevertheless, by studying many different initial configurations, we have observed that for (small) negative values of $\lGB$, the isotropisation time is always larger than for $\lGB=0$, while  the opposite is true for positive $\lGB$ values. In the linearised regime of moderately large anisotropies, we demonstrate that this correlation of the change of isotropisation time with the sign of $\lGB$ always holds, irrespective of initial conditions. Quite satisfactorily, Gauss-Bonnet gravities with negative (positive) $\lGB$ values are dual to gauge theories with shear viscosity to entropy density ratio $\eta/s$, larger (smaller) than $1/4\pi$ \cite{Myers-GB}, the value of the ratio for infinitely strongly coupled SYM. This correlation further supports the interpretation that Gauss-Bonnet gravity represents a good tool to understand the finite coupling corrections of $\N=4$ SYM.

This paper is organised as follows: in Section~\ref{preliminaries} we will review the isotropisation of large initial configurations in $\N=4$ SYM, and also gather the key ingredients of Gauss-Bonnet gravity that we will need for our analysis. In Section~\ref{sec:time evolution GB} we introduce the numerical procedure we employ to study small $\lGB$ corrections to the isotropisation process, and discuss the main systematics of our numerical results over many different initial configurations. In Section~\ref{sec:QNM} we focus on the effective linear  regime, and show how the systematics observed in Section~\ref{sec:time evolution GB}  follow from the analysis of the associated quasinormal mode expansion. 
In Section~\ref{sec:discussion} we perform a systematic exploration over initial conditions to determine the validity of our findings in the effectively linear regime.  
Finally, in Section~\ref{sec:conclusions} we summarise our main results and discuss their implications for finite coupling corrections of holographic theories.


\section{Preliminaries}
\label{preliminaries}

\subsection{Isotropisation in strongly coupled $\N=4$ SYM}

In this  section we briefly review the procedure to study the isotropisation of far-from-equilibrium initial states in strongly coupled $\N=4$ SYM, following \cite{Chesler:2008hg, HoloLinPRL,HoloLinJHEP,CYReview}. This section will also serve to establish the notation and the general strategy to solve the isotropisation problem with higher derivative corrections, which we will perform in the next section. 

We prepare an anisotropic yet homogeneous state of $\N=4$ by specifying an initial value for the bulk metric in the gravity dual. We will parametrise the (4+1)-dimensional space dual to the field theory state by the following metric ansatz: 
\eqn{BulkAnsatz}
{ds^2=-2A_0 dt^2 + \Sigma_0^2 \left(e^{B_0} dx_1^2 + e^{B_0} dx_2^2 + e^{-2B_0} dx_3^2\right) + 2dtdr\,,}
\noindent where $A_0$, $\Sigma_0$ and $B_0$ are all functions of $t$ and $r$ only, and the boundary is at $r\to\infty$.  This ansatz enjoys and residual gauge freedom which arises from reparametrizations of the holographic coordinate, $r\rightarrow r+\lambda(t)$.
 Introducing this ansatz into Einstein's equations and imposing that the space is asymptotically AdS, the near boundary expansion 
 of the different metric fields is given by 
\eqn{BC}
{\begin{aligned}
A_0 &= \frac{1}{2 L^2}(r+\lambda)^2-\partial_t\lambda+\frac{L^6 a_0^{(4)}}{r^2}+\mathcal{O}\left(r^{-3}\right)\,,\cr
\Sigma_0 &= \frac{1}{L}(r+\lambda) +\mathcal{O}\left(r^{-7}\right)\,,\cr
B_0 &= \frac{L^8 \hat g_{0,11}^{(4)}}{r^4} +\mathcal{O}\left(r^{-5}\right)\,,\cr
\end{aligned}}
where $\lambda(t)$ is arbitrary, and $\hat g_{0,11}^{(4)}$ and $a_0^{(4)}$ are unknown functions of time which cannot be determined from a power series expansion. 
Here $L$ is the AdS radius defined in terms of the cosmological constant $\Lambda$ via
\begin{equation}
	\Lambda = - 6/L^2\,.
\end{equation}
The asymptotic expansion \eqref{BC} determines the stress tensor of the dual gauge theory via holographic renormalisation \cite{Bianchi:2001de,Bianchi:2001kw} to be
\eqn{DualStressExplN4}
{\hat T_{ab}= \,{\rm diag}\left(-\frac{3}{2}a_0^{(4)},\,\,\,\,\hat g_{0,11}^{(4)}-\frac{1}{2}a^{(4)},\,\,\,\,\hat g_{0,11}^{(4)}-\frac{1}{2}a_0^{(4)},\,\,\,\, -2\hat g_{0,11}^{(4)}-\frac{1}{2}a_0^{(4)}\right)\,,}
with $  T_{ab}=  2 \hat T_{ab} L^3/\kappa^2_5 \, $ and $\kappa^2_5$ the five dimensional Newton constant, which is related to the number of colours of the dual gauge theory by $L^3/\kappa^2_5=N_c^2/4 \pi^2$.
We work in units of $L=1$ henceforth.

With the gauge choice in \eno{BulkAnsatz}, Einstein's equations contain only 5 non-vanishing components. After defining the modified time derivative as
\eqn{dPlusDef}
{d_+\equiv \partial_t + A_0\,\partial_r\,,}
Einstein's equations take the following nested form
\begin{eqnarray}\label{NestedEoM1}
0 &=& \Sigma_0''+\frac{1}{2}\Sigma_0 \, (B_0')^2\,,\\ \label{NestedEoM2}
0 &=& (d_+\Sigma_0)'+2(d_+\Sigma_0)\frac{\Sigma_0'}{\Sigma_0}-2\Sigma_0\,, \\ \label{NestedEoM3}
0 &=& (d_+B_0)'\,\Sigma_0+\frac{3}{2}(d_+B_0)\Sigma_0'+\frac{3}{2}(d_+\Sigma_0) B_0'\,, \\ \label{NestedEoM4}
0 &=& A_0'' -6(d_+\Sigma_0)\frac{\Sigma_0'}{\Sigma_0^2}+\frac{3}{2}(d_+B_0) B_0' + 2\,,  \\ \label{NestedEoM5}
0 &=& d_+ d_+ \Sigma + \frac{1}{2} (d_+ B)^2 \Sigma - A' d_+ \Sigma \,,
\end{eqnarray}
where primes denote $r-$derivatives. 
One can check that \eno{NestedEoM5} yields $d a_0^{(4)} (t)/dt=0$, which, via \eno{DualStressExplN4}, implies that the energy density of the system is conserved. Otherwise, it does not participate in the dynamics and can be dropped as long as $a_0^{(4)}$ is held constant.
This nested form provides a convenient setup in which 
the metric functions $\Sigma_0$, $d_+\Sigma_0$, $d_+B_0$ and $A_0$ can be determined sequentially at every time slice by solving linear
ODE's, once the $r$-dependence of $B_0$ is known. 
Therefore, the time evolution of the metric functions can be obtained after determining $\partial_t B_0$ from \eno{dPlusDef} and knowledge of $d_+B_0$ and $A_0$ at each time slice. 
 The above nested pattern allows us to specify a full set of 
 initial states of the dual gauge theory by specifying the functional form of the metric field  $B_0(t_0,r)$  at some initial time $t_0$.  These states provide a convenient framework in which to study far-from-equilibrium dynamics with a variety of initial conditions.
  
To solve the dynamical equations \eqref{NestedEoM1}-\eqref{NestedEoM4}, it is  convenient to redefine the metric fields to facilitate the imposition of the boundary conditions \eqref{BC}.  
Following \cite{CYReview}, after the coordinate transformation $u\equiv 1/r$, we define:
\eqn{Redef}
{\begin{aligned}
A_0 &\equiv \frac{1}{2}\left(\frac{1}{u} + \lambda \right)^2 + a_0\, , \\ 
B_0 &\equiv u^3\, b_0\,, \quad \\ 
\Sigma_0 &\equiv \frac{1}{u} + \lambda +u^4\,\sigma_0\,, 
\cr
d_+ \Sigma_0 &\equiv \frac{1}{2}\left(\frac{1}{u} + \lambda  \right)^2+u^2\,\dot \sigma_0\,,  \quad \\ 
d_+B_0 &\equiv u^3\, \dot b_0\,. \quad
&
\end{aligned}}
Note that in these redefinitions the functions $\dot \sigma$ and $\dot b$ are $not$ the time derivatives of $\sigma$ and $b$, but rather independent functions in the same way in which 
 the metric functions $d_+B$ and $d_+\Sigma$ are independent from $B$ and $\Sigma$ at each time slice in the nested procedure to solve Einstein's equation. 
Imposing  \eno{BC} and the boundary expansion of $d_+ \Sigma_0$ and $d_+B_0$, these redefined fields satisfy the following boundary conditions in the $u\rightarrow 0$ limit:
\eqn{RedefBC}
{\sigma_0\to u^3 ,\quad b_0\to u\, \hat g_{0,11}^{(4)}\,,\quad a_0\to -\partial_t\lambda + u^2\,a_0^{(4)}\,,\quad \dot \sigma_0 \to a_0^{(4)}\,,\quad \dot b_0\to -2\hat g_{0,11}^{(4)}\,.}

For a given gauge choice $\lambda(t)$, the nested system of equations \eqref{NestedEoM1} together with the redefinitions \eqref{Redef} and the boundary conditions \eqref{RedefBC}, fully specify the time evolution of the system once the initial data 
 $b_{0, \, \rm init} =b_0(0,u)$ is specified.
For the computations of this work, in the $\lambda=0$ gauge, we consider manny different arbitrary initial data $b_{0, \, \rm init} $.
 These are constructed as the ratio of two 10th order polynomials in $u$ whose coefficients are chosen randomly in the range [0,1]. We will 
furthermore multiply this ratio by a random amplitude (in the range [0,10]), and also subtract the constant term, so that the boundary conditions \eqref{RedefBC} are obeyed\footnote{This procedure can, in some cases, generate caustics. In those cases, the amplitude of the initial condition is gradually reduced until the caustics appear behind the apparent horizon.}. We have also tested the Gaussian initial conditions studied in \cite{CYReview} with varying amplitude. 


At each time slice, we solve for the metric components via pseudospectral methods on the Chebyshev grid in the holographic coordinate $u$. Since generic choices of $b_{0, \, \rm init} $ will lead to the formation of an apparent horizon in the bulk metric \cite{HoloLinJHEP}, we will choose our grid to be $u\in[0,u_{\rm H}]$, where $u_{\rm H}$ is the location of the apparent horizon, given by the condition
\eqn{AppHor}
{d_+\Sigma_0(t,u_H)=0\,.}
In the $\lambda=0$ gauge this location changes with time which complicates the application of pseudospetral methods. For this reason, after locating the apparent horizon in the initial time slice, we reparametrise the holographic coordinate such that the position of the horizon is fixed. Without loss of generality we choose $\lambda$ such that $u_{\rm H}=1$ in our numerics. This, in turn, implies that $\lambda$ becomes a dynamical variable that must be updated at each step on the time integration. The time derivative of $\lambda$ may be found via the horizon stationarity condition \cite{CYReview}, i.e. the time derivative of \eno{AppHor},  which, after using the equations of motion, boils down to\footnote{Note that the apparent horizon condition \eqref{AppHor} implies that the right hand side of this equation vanishes identically at the apparent horizon. Nevertheless, the numerical procedure used to determine the initial apparent horizon implies that the left hand side of \eno{AppHor} is never identically zero. We have found that keeping explicitly the right hand side of \eno{HorStat} improves the stability of our code significantly.}
\eqn{HorStat}
{A_0+\frac{1}{4}(d_+B_0)^2=\frac{d_+\Sigma_0}{2\Sigma_0}\left(A_0'+\frac{2A_0\Sigma_0'}{\Sigma_0}\right)\,.}
To obtain $\partial_t \lambda$ at a given time slice, we solve the first three nested equations \eqref{NestedEoM1}-\eqref{NestedEoM3} using the explicit boundary conditions \eqref{RedefBC}. After solving these equations, \eno{HorStat} may be viewed as imposing a boundary condition for \eno{NestedEoM4} at the apparent horizon. We can then solve for $a_0(t,u)$ and from \eno{RedefBC}, $\partial_t \lambda$ is then given by$-a_0(t,0)$. 
Combining this derivative with $d_+ B_0$, obtained from \eno{NestedEoM3} and  with the redefinition of the fields \eno{Redef}, we can determine
 $\partial_t b_0$. This allows us to determine $b_0$ at the next time slice and, subsequently, all other metric functions. Iterating the procedure at every time step determines the metric at all times. We perform the time evolution using a fourth order Runge-Kutta algorithm. 
 
From the evolution of the metric we can extract the stress tensor of the anisotropic state in the dual field theory at later times. Given that the energy density remains constant throughout the evolution and that the trace of the stress tensor vanishes identically, the only non-trivial component is  the pressure anisotropy, $\Delta p_0\equiv T_{0,zz}-(T_{0,xx}+T_{0,yy})/2$. Combining Eqs.
\eqref{DualStressExplN4}, \eqref{Redef} and \eqref{RedefBC}, the pressure anisotropy is given by 
\eqn{PressureExplicit}
{
\Delta \hat p_0=-3(\partial_u b_0)|_{u=0}\,,
}
where the hatted quantities have the same definition as in \eno{DualStressExplN4}. For very anisotropic initial estates, the pressure anisotropy $\Delta p_{0}$ can be large at initial time $t_0$. As time passes, the anisotropy decays until, at sufficiently late time the geometry becomes that of the 
AdS-Schwarzschild black hole, given by \eno{BulkAnsatz} with 
\begin{equation}
	B_0 = 0 , \qquad \Sigma_0 = r , \qquad A_0 = \frac{r^2}{2}\left( 1 - \frac{r_H^4}{r^4} \right)  \,.
\end{equation}

In terms of the field theory dual, the stress tensor relaxes to the equilibrium stress tensor in which all pressures 
are the same and equal to $p_{\rm eq}=\epsilon /3$, with $\epsilon$ the energy density which is determined by $a_0^{(4)}$ as
$\epsilon=-\frac{3}{4}\frac{N_c^2}{\pi^2}a_0^{(4)}$ via \eno{DualStressExplN4}.
At these late times, the system is in equilibrium, with a temperature $T_0$ related to the energy density via the equation of state of $\N=4$ SYM, $\epsilon= N_c^2 \hat \epsilon /4\pi$ with
\begin{equation}
\label{ET0}
\hat \epsilon= \frac{3}{4} \pi^4 T_0^4 \,.
\end{equation}
 This relaxation process is called isotropisation and has been studied in detailed in previous works 
 \cite{Chesler:2008hg, HoloLinPRL,HoloLinJHEP,CYReview}. These studies constitute the basis we will employ to study isotropisation in Gauss-Bonnet theories. 

%
\subsection{Gauss-Bonnet gravity}

In this section we collect some useful results about Gauss-Bonnet gravity with a 
negative cosmological constant \cite{Cai-GBSoln}.
The five-dimensional action we consider takes the form
\eqn{GBAction} 
{S_{GB} = \frac{1}{2 \kappa^2_5}\int d^5 x\sqrt{-g}\left(R + 12 +
\frac{\lambda_{GB}}{2}\left( R_{\mu\nu\rho\sigma}R^{\mu\nu\rho\sigma} -4R_{\mu\nu}R^{\mu\nu}+R^2\right) \right)\,,}
where $\lambda_{GB}$ is a dimensionless number, constrained (at least for holographic purposes) by causality \cite{Myers-GB} and positive definiteness of the boundary energy density \cite{Buchel:2009tt} to be\footnote{See also \cite{Andrade:2016yzc} for a recent revision 
of this bound which takes into account hyperbolicity considerations.} 
\eqn{LambdaGB}
{-\frac{7}{36}<\lambda_{GB}\leq \frac{9}{100}\,.}
Einstein equations coming from \eno{GBAction} can be written as
\eqn{EEGB}
{E_0+\lambda_{GB}E_{GB}=0\,,}
\noindent where $E_0$ are the zeroth order ($\lambda_{GB}=0$) Einstein equations
\begin{equation}
	E_{0, \mu \nu}(g) = G_{\mu \nu} + \Lambda g_{\mu \nu}\,,
\end{equation}
\noindent and the contribution from the Gauss-Bonnet term is given by
\begin{align}
\nonumber
	E_{GB, {\mu \nu} }(g)= &   R R_{\mu \nu} +  R_{\mu \alpha \beta \gamma} R_\mu \hs ^{\alpha \beta \gamma} 
	- 2 R_{\mu \alpha} R^\alpha_\nu - 2 R^{\alpha \beta}  R_{\alpha \mu \beta \nu} \\
	&- \frac{1}{4} g_{\mu \nu} \left(  R_{\alpha \beta \gamma \delta} R^{\alpha \beta \gamma \delta} 
	- 4 R_{\alpha \beta} R^{\alpha \beta} + R^2 \right)\,.
\end{align}
The near boundary behaviour of the metric fields can be easily found to be
\begin{align}\label{bc for GB}
A &= \frac{1}{2 L_c^2}(r+\lambda)^2-\partial_t\lambda+\frac{a_0^{(4)}}{r^2}+\mathcal{O}\left(r^{-3}\right)\,,\cr
\Sigma &= \frac{1}{L_c}(r+\lambda) +\mathcal{O}\left(r^{-7}\right)\,,\cr
B &= \frac{\hat g_{0,11}^{(4)}}{r^4} +\mathcal{O}\left(r^{-5}\right)\,,
\end{align}
\noindent where
\eqn{EffL}
{L_c= \sqrt \frac{1+U}{2}\,,\qquad U\equiv \sqrt{1-4\lambda_{GB}}\,.}
This follows from the fact that the solutions of \eno{GBAction} are asymptotically AdS with the effective radius $L_c$ \cite{GBHoloRenorm}, 
and can be of course checked explicitly. 

The holographic renormalisation for Gauss-Bonnet gravity has been performed in \cite{GBHoloRenorm, Astefanesei:2008wz}, where the covariant counterterms
were computed. The resulting boundary stress tensor turns out to be
\eqn{BndStress}
{\mathcal{T}_{ab}=\frac{1}{2} \left( K_{ab}-\gamma_{ab}K+\lambda_{GB}\left(Q_{ab}-\frac{1}{3}Q \gamma_{ab}\right)-\frac{2+U}{L_c}\gamma_{ab}+\frac{L_c}{2}(2-U)\left( \mathcal{R}_{ab}-\frac{1}{2}\gamma_{ab}\mathcal{R} \right)\right)\,,}
where Roman indices go over the boundary directions, $\gamma_{ab}$ is the induced metric on the boundary, $K_{ab}$ is the extrinsic curvature of the boundary (and $K$ is its trace), and $Q_{ab}$ is a tensor (whose explicit form can be found in \cite{GBHoloRenorm}) given in terms of the extrinsic curvature, the Ricci scalar $\mathcal{R}$, Ricci tensor $\mathcal{R}_{ab}$ and the Riemann tensor $\mathcal{R}_{abcd}$ associated with the boundary metric $\gamma_{ab}$.

The expectation value $\hat T_{ab}$ of the dual theory stress tensor is then given by 
\eqn{DualStress}
{\sqrt{-h}\,h^{ab}\hat T_{bc}=\lim\limits_{r\to\infty}\sqrt{-\gamma}\,\gamma^{ab}\mathcal{T}_{bc}\,,}
where hat denotes the same rescaled definition of the stress tensor as in \eno{DualStressExplN4}, and where $h_{ab}$ is the background metric on which the dual theory lives, given by 
\eqn{DualMetric}
{h_{ab}=\lim\limits_{r\to\infty}\frac{L_c^2}{r^2}\gamma_{ab}\,,}
which, using the boundary conditions \eqref{BC} with $L\to L_c$, evaluates to ${\rm diag} (-1,1,1,1)$, as it should. In the end, we get:
\eqn{DualStressExpl}
{\hat T_{ab}=\frac{2L_c^2-1}{L_c^3}\,{\rm diag}\left(-\frac{3}{2}a^{(4)},\,\,\,\,\frac{\hat g_{11}^{(4)}}{L_c^2}-\frac{1}{2}a^{(4)},\,\,\,\,\frac{\hat g_{11}^{(4)}}{L_c^2}-\frac{1}{2}a^{(4)},\,\,\,\, -2\frac{\hat g_{11}^{(4)}}{L_c^2}-\frac{1}{2}a^{(4)}\right)\,.}
Note that setting $\lambda_{GB}= 0$ or equivalently $L_c=1$, we arrive at the same expression as in \eno{DualStressExplN4}.

Similarly to the asymptotically AdS case, we expect the final state of the isotropisation in Gauss-Bonnet to be the corresponding black 
brane solution in this theory. The line element takes the form \eqref{BulkAnsatz} with \cite{Cai-GBSoln}\footnote{See also 
\cite{Boulware:1985wk} for an earlier reference which found this solution with vanishing cosmological constant.}
\begin{equation}\label{GB bh}
	B = 0 , \qquad \Sigma = \frac{r}{L_c} , \qquad A = \frac{r^2}{4 \lambda_{GB}} 
	\left \{ 1 - \left[1 - 4 \lambda_{GB} \left( 1 - \frac{r_H^4}{r^4} \right) \right]^{1/2}   \right \}
\end{equation}
The event horizon is located at $r = r_H$. The choice of the minus sign in front of the square root in \eno{GB bh}
makes the limit $\lambda_{GB} = 0$ well defined so that this geometry is smoothly connected to that of the 
AdS-Schwarzschild black hole.\footnote{In addition, the choice of the plus sign leads to ghosts \cite{Cai-GBSoln}.} Note that the boundary speed of light associated with \eno{GB bh} is unity. 
The Hawking temperature of this solution is given by
\begin{equation}
	T = \frac{r_H}{\pi}\,,
\end{equation}
and the energy density, as given by $\hat \epsilon = \hat T_{00}$ in \eno{DualStressExpl}, turns out to be
\begin{equation}
\label{ETlambda}
	\hat \epsilon =  \frac{3 \pi^4}{4 L_c^3} T^4\,.
\end{equation}

\section{Time evolution in Gauss-Bonnet gravity}
\label{sec:time evolution GB}

Our strategy to study time evolution in Gauss-Bonnet gravity is to linearise the problem in the coupling $\lambda_{GB}$.
In addition to the obvious technical advantage that the evolution equations are linear, this approach ensures that the 
dynamics remain hyperbolic, so that we can evolve in time for a given initial condition specified on a null slice, as we shall see explicitly below. For finite values of $\lambda_{GB}$, characteristic surfaces -- i.e. the places on which we wish to specify initial data --
in general do not coincide with metric null cones, which in certain cases leads to the initial value problem being ill-defined
\cite{Reall:2014pwa, Andrade:2016yzc}\footnote{We thank Pau Figueras for bringing up this point.}. Working in perturbation theory around $\lambda_{GB} =0$ allows us to overcome this difficulty because the dynamical structure of the perturbations is that of the underlying Einstein-Hilbert equations of motion, thus, much of the 
formalism of \cite{Chesler:2008hg} which we employ to determine the background solution carries over to the Gauss-Bonnet case.

\subsection{Linearised time evolution}

Having found a (time-dependent) background solution $g_0(t,r)$ of the non-linear equations $E_0$, we wish to linearise 
Einstein equations around it by postulating a solution of the form
\eqn{LinRule}
{g=g_0+\lambda_{GB}\,\delta g\,.}
Plugging this into \eno{EEGB} and keeping terms up to first order in $\lambda_{GB}$, we get
\eqn{EELin}
{E_{\rm 0, lin}[\delta g]=-E_{GB}[g_0]\,,}
where $E_{\rm 0, lin}$ are the zeroth order Einstein equations linearised around the background solution $g_0(t,r)$, for which 
the Gauss-Bonnet contribution $E_{GB}$ evaluated on the background acts as a source. 
In addition to the terms determined explicitly by the background evolution, the source receives contributions 
which involve $d_+d_+B_0$, $d_+d_+\Sigma_0$ and $d_+A_0$. While $d_+d_+\Sigma_0$ can be determined from the zeroth order equations $E_0$, these do not contain $d_+d_+B_0$ nor $d_+A_0$. This does not present a problem because we can obtain these extra terms
(or simply $\partial_t^2 B_0$ and $\partial_t A_0$ if one wishes) numerically since we know the background solution for all $t$ and $r$. 

The form of the resulting linear equations \eqref{EELin} makes it clear that the dynamics of the perturbations $\delta g$ inherit the structure 
of the background equations of motion so that, in particular, we can cast the evolution problem as a set of nested 
ODE's.\footnote{The equations that we need to solve are linear PDE's in the variables $(t,r)$, so our approach 
provides a great technical advantage of reducing the problem to linear ODE's. The fact that these are nested is however not crucial due to linearity.}
To do so, it suffices to appropriately linearise the definition of $d_+$ by letting, for any perturbation $\delta F$, 
\eqn{HattedDPlus}
{\hat d_+ \delta F\equiv \partial_t\delta F+A_0\partial_r \delta F + \delta A \partial_r F_0\,.}
The steps to solve for the linear evolution problem then closely parallel those of the background: after specifying
the initial condition for the metric anisotropy on an initial time slice and the boundary condition that fixes the perturbed 
energy density, 
we solve the set of ODE's for $\delta \Sigma$, $\hat d_+ \delta \Sigma$, $\hat d_+ \delta B$
and $\delta A$, which allows us to calculate $\partial_t \delta B$ and time evolve to the next slice. 
%
%
Notice that the small perturbation problem we are considering can be formulated in the same coordinates as the background evolution and no additional $\lGB$-dependent shift in the gauge parameter $\lambda(t)$ needs to be performed. This is a consequence of demanding regularity of the solution at the event horizon, which, in the Eddington Finkelstein coordinates we use, implies in-falling boundary conditions for the perturbation. Because of this regularity, infinitesimally small shifts in the position of the event horizon do not lead to information loss in the boundary theory.
We should also note that, since we are consistently linearising in $\lambda_{GB}$, we do not have to specify its value at any step in the algorithm, since this parameter appears outside of $\delta g$ (\eno{LinRule}), and hence the solution $\delta B(t,u)$ is $\lambda_{GB}$-independent. 

In order to impose the boundary conditions, we record the asymptotics satisfied by the linearised fields, which directly follow from 
\eno{bc for GB} 
\eqn{PertBC}
{\begin{aligned}
\delta A &= \frac{1}{2}(r+\lambda)^2 + \frac{\delta a^{(4)}}{r^2}+ \mathcal{O}(r^{-3})\,,\cr
\delta \Sigma &= \frac{1}{2}(r+\lambda)+ \mathcal{O}(r^{-7})\,,\cr
\delta B &= \frac{\delta \hat g_{11}^{(4)}}{r^4}+\mathcal{O}(r^{-5})\,,
\end{aligned}}
\noindent after choosing 
\eqn{a4b4}
{a^{(4)}=a_0^{(4)}+\lambda_{GB}\delta a^{(4)}\,,\qquad \hat g_{11}^{(4)}=\hat g_{0,11}^{(4)}+\lambda_{GB}\delta \hat g_{11}^{(4)}\,.}
Thus, requiring a set of boundary conditions in the UV reduces to specifying the values of the linearised contributions 
to the stress tensor $\delta \hat g_{11}^{(4)}$ and $\delta a^{(4)}$. 
In solving the linear ODE's for the perturbations, we find it convenient to introduce the redefinitions
\eqn{RedefPert}
{\begin{aligned}
\delta\Sigma &\equiv \frac{1}{2}\left(\frac{1}{u} + \lambda\right)+u^4\,\delta\sigma\,,\cr
\hat d_+ \delta\Sigma &\equiv \frac{3}{4}\left(\frac{1}{u} + \lambda  \right)^2+u^2\,\delta\dot \sigma\,,\cr
\delta A &\equiv \frac{1}{2}\left(\frac{1}{u} + \lambda  \right)^2 + \delta a\,,\cr
\delta B &\equiv u^3\, \delta b\,,\cr
\hat d_+\delta B &\equiv u^3\, \delta\dot b\,,
\end{aligned}}
\noindent which resemble those of \eno{Redef}. Notice that, as in \eno{Redef}, $\delta \dot \sigma$ and $\delta \dot b$ are not the time derivatives of $\delta \sigma$ and $\delta b$, but independent fields.
The boundary conditions as $u \to 0$ for the redefined fields are 
\eqn{RedefPertBC}
{\delta\sigma\to 0\,,\quad \delta \dot \sigma \to \frac{a_0^{(4)}}{2}+\delta a^{(4)} \,,\quad \delta a \to u^2 \delta a^{(4)}\,,\quad \delta b\to u \,\delta \hat g_{11}^{(4)}\,,\quad \delta \dot b \to -2 (\hat g^{(4)}_{0,11}+\delta \hat g_{11}^{(4)})\,.}

Since we want to study the change in isotropisation between the same state at different values of $\lGB$,
we impose boundary conditions for the perturbations such that the energy density and initial pressure anisotropy of the Gauss-Bonnet solution coincide with that of the $\lambda_{GB} = 0$ solution. From the expression for the stress tensor, 
equating the energy densities at zero and non-zero $\lambda_{GB}$, taking into account \eno{a4b4} and linearising 
in $\lambda_{GB}$,  we arrive at
\begin{equation}\label{DeltaA4}
	\delta a^{(4)}  = \frac{1}{2} a^{(4)}_0\,.
\end{equation}
Similarly, up to linear order in $\lambda_{GB}$, the pressure anisotropy is given by
\begin{align}\label{DeltaDeltaP}
\Delta \hat p &= -3(\partial_u b_0)_{u=0} - 3\lambda_{GB}\left( (\partial_u \delta b)_{u=0} + \frac{1}{2}(\partial_u b_0)_{u=0}\right)+\mathcal{O}\left(\lambda_{GB}^2\right)\cr
&\equiv  \Delta \hat p_0 + \lambda_{GB} \delta(\Delta \hat p)\,,
\end{align}
\noindent where we have defined the perturbed pressure anisotropy
\begin{equation}\label{dDp}
	\delta(\Delta \hat p) = - 3 \left( (\partial_u \delta b)_{u=0} + \frac{1}{2}(\partial_u b_0)_{u=0}\right) \,.
\end{equation}

The requirement that the initial expectation value of the stress tensor is the same  independently of the value of $\lGB$ does not completely
fix the state since it only constraints its near-boundary behaviour. However, our approach requires to specify the perturbation 
$\delta b(u,t=0)$ for all $u$. The simplest possible initial choice consistent with the former requirement is that
 \eno{dDp} vanishes at all $u$, i.e.
\begin{equation}\label{db IC}
	\delta b(u,t = 0) = - \frac{1}{2} b_0(u, t=0) \,.
\end{equation}
All the numerical results presented in this Section will be performed with this initial condition.  In Sec. \ref{sec:QNM} we will come back to this point and discover that for a large class of initial perturbations, the isotropisation dynamics depend
very weakly on the particular choice of initial condition for $\delta b$.

\begin{figure*}[ht!]
\centering
\includegraphics[height=0.29\textwidth]{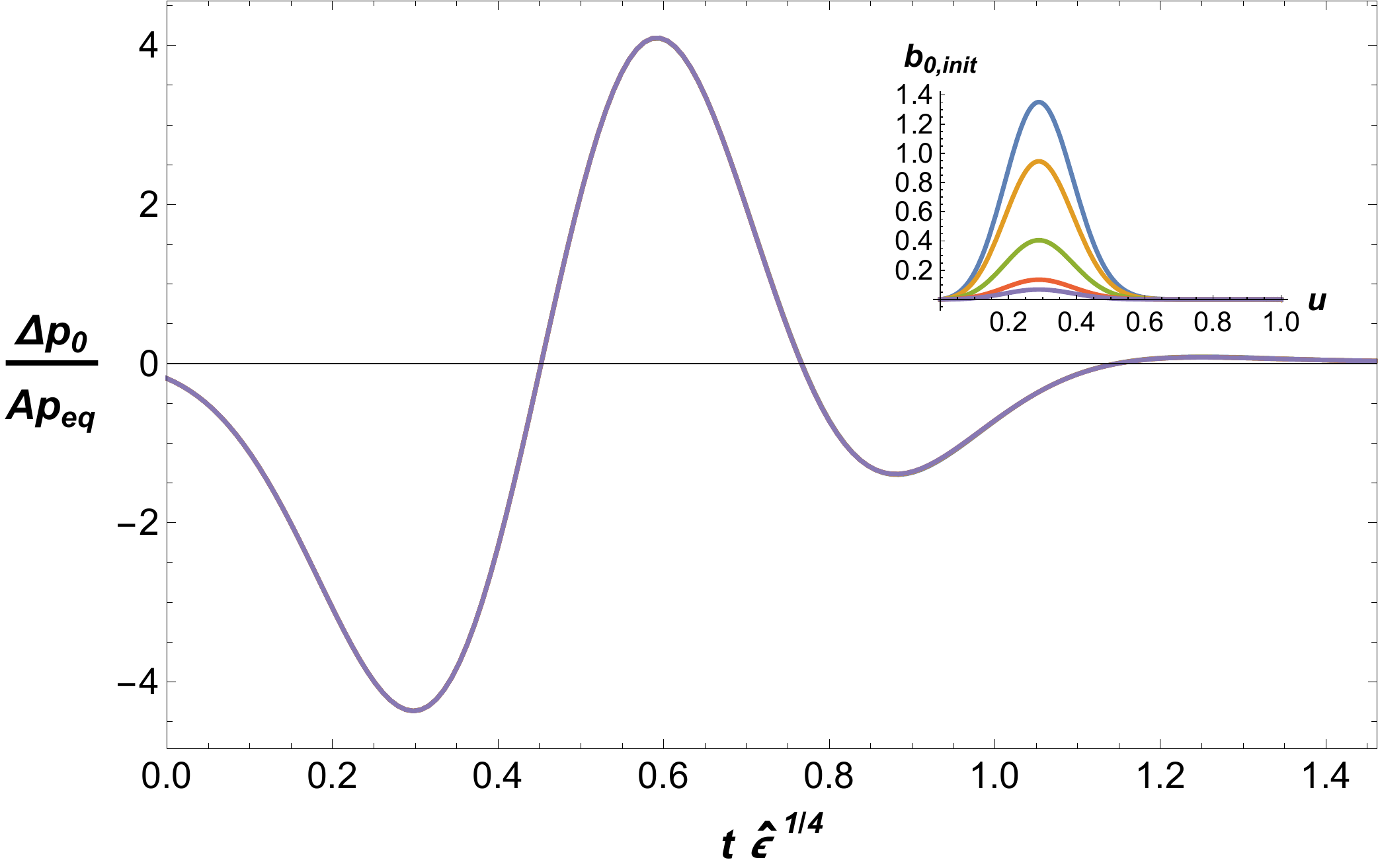}
\hskip 0.02\textwidth
\includegraphics[height=0.29\textwidth]{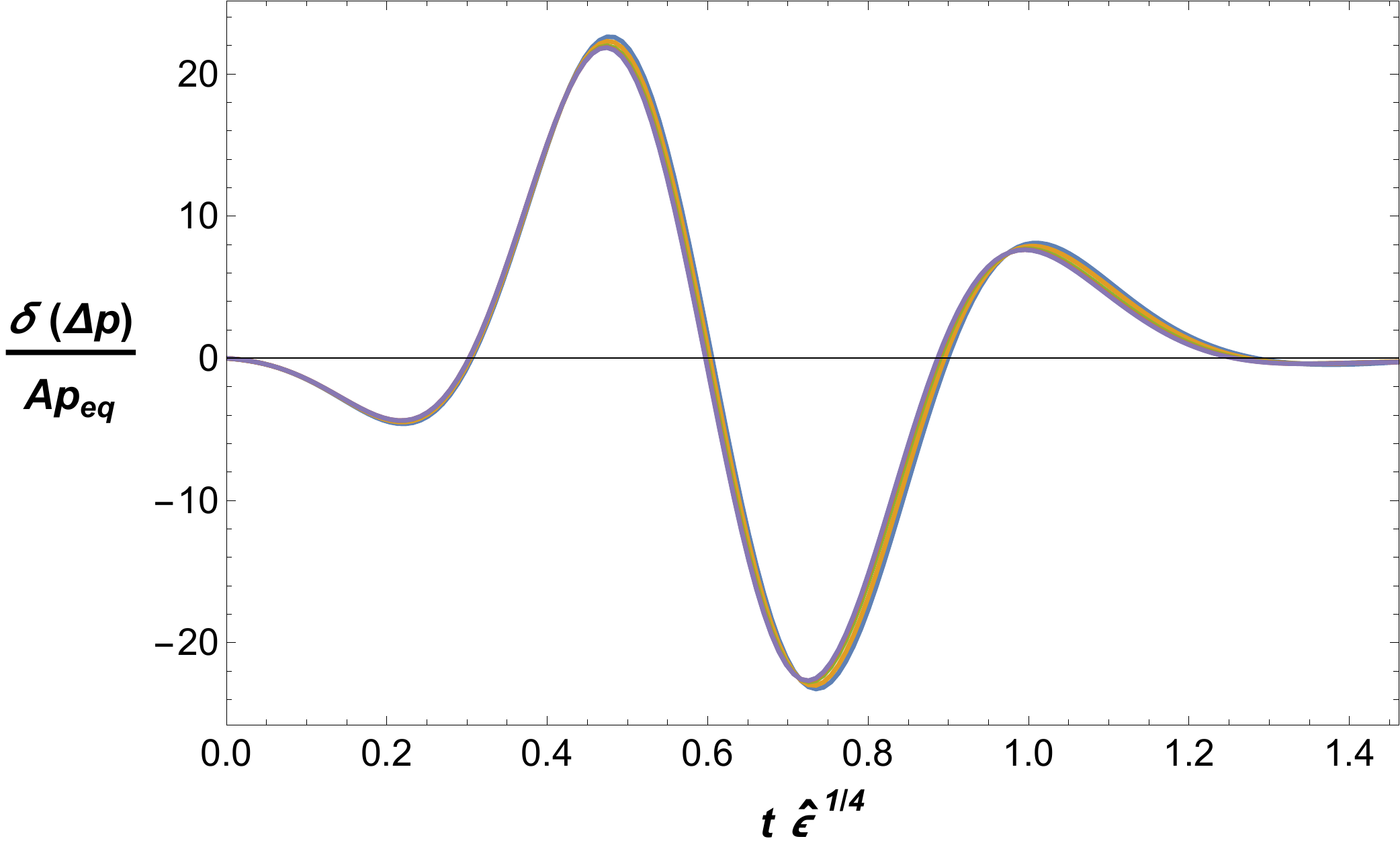}
\vskip 0.01\textwidth
\includegraphics[height=0.29\textwidth]{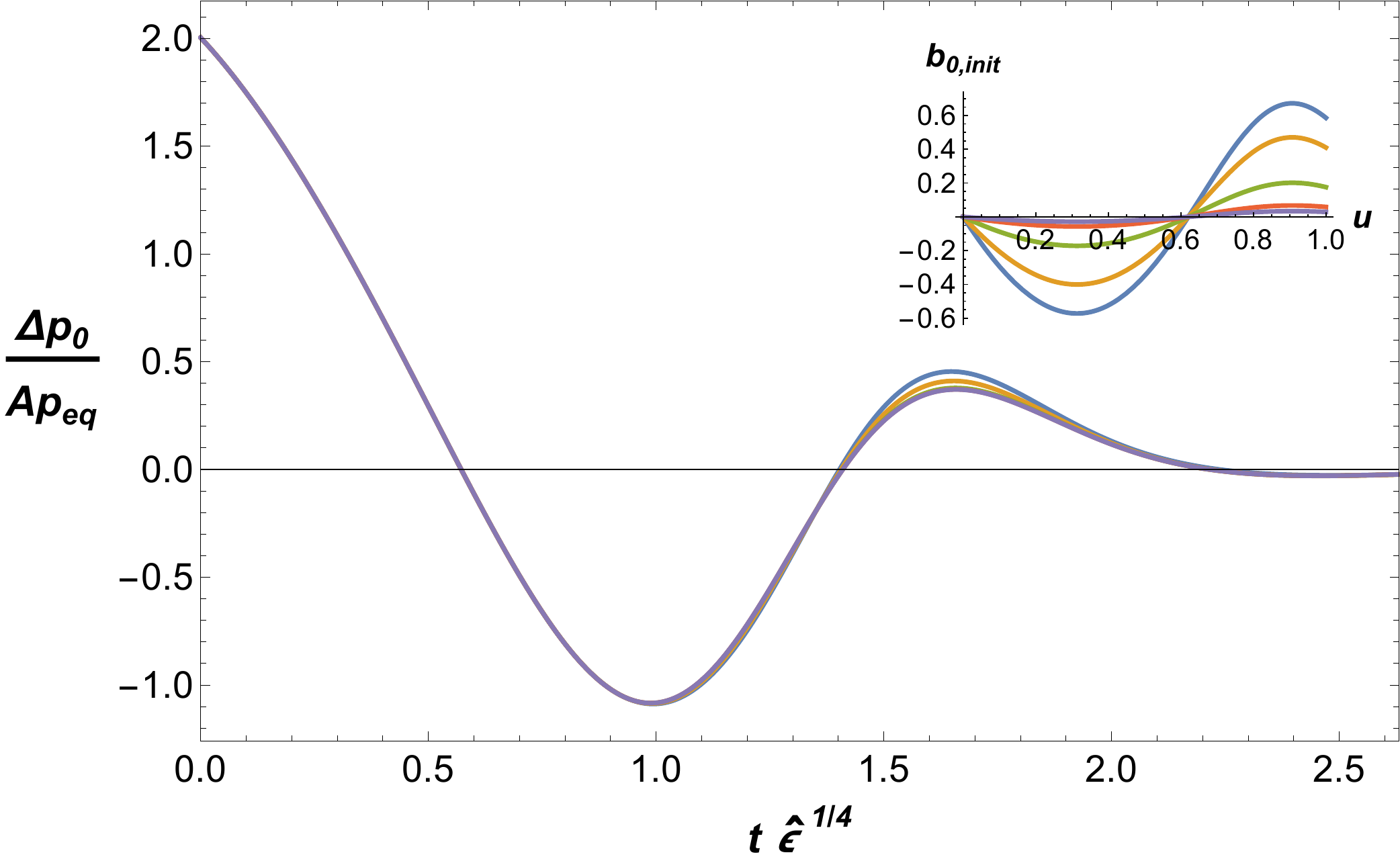}
\hskip 0.02\textwidth
\includegraphics[height=0.29\textwidth]{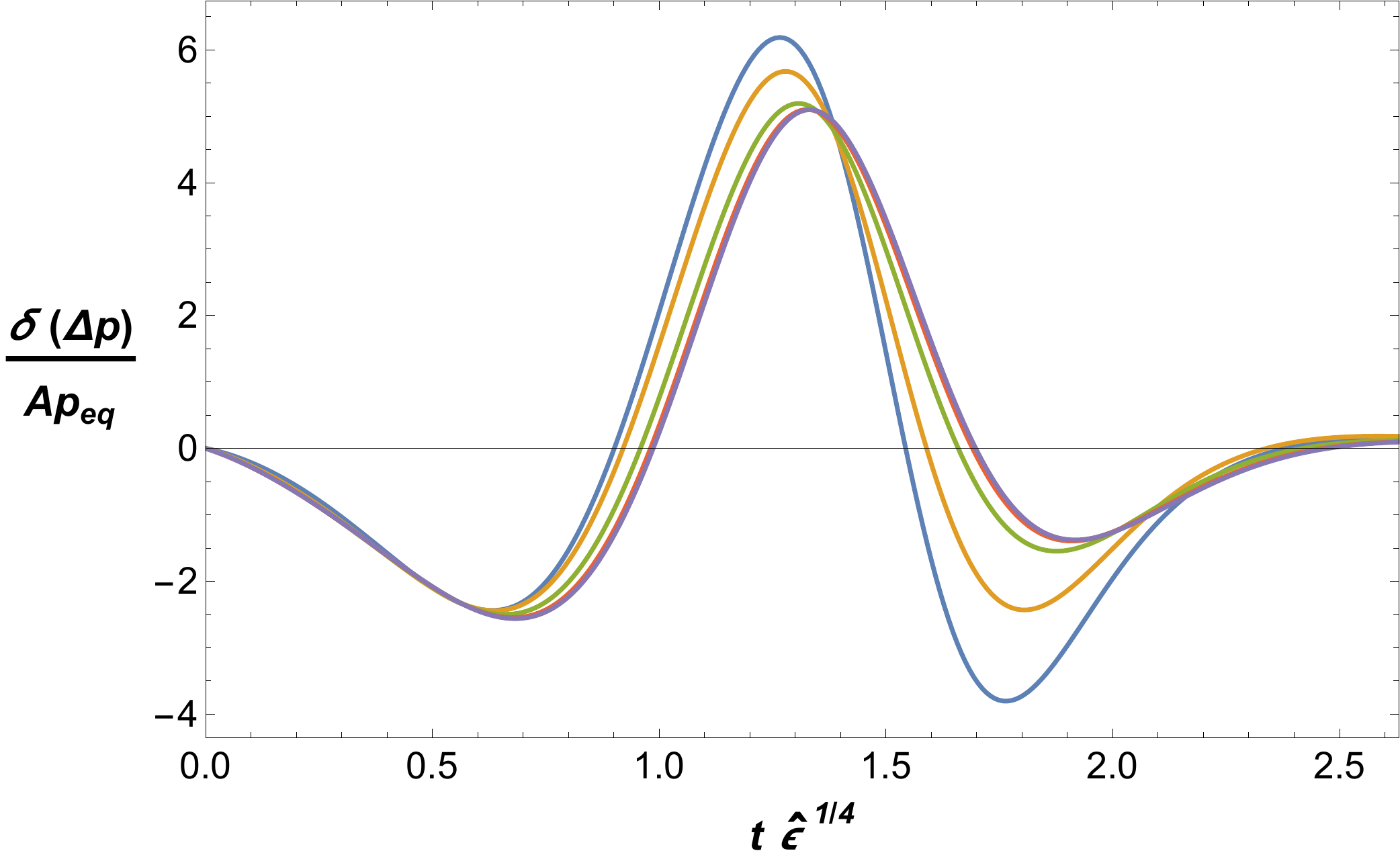}
\vskip 0.01\textwidth
\includegraphics[height=0.29\textwidth]{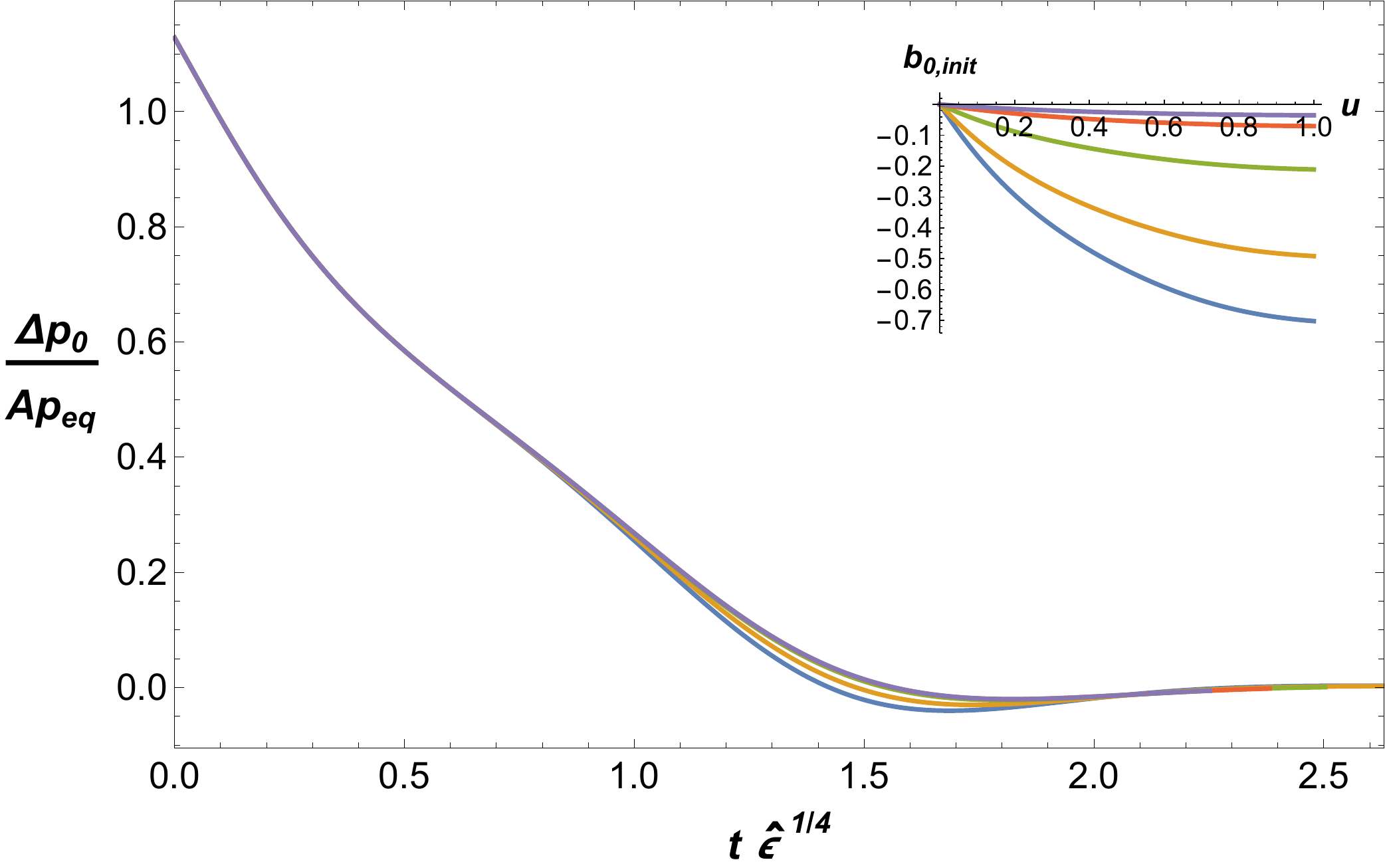}
\hskip 0.02\textwidth
\includegraphics[height=0.29\textwidth]{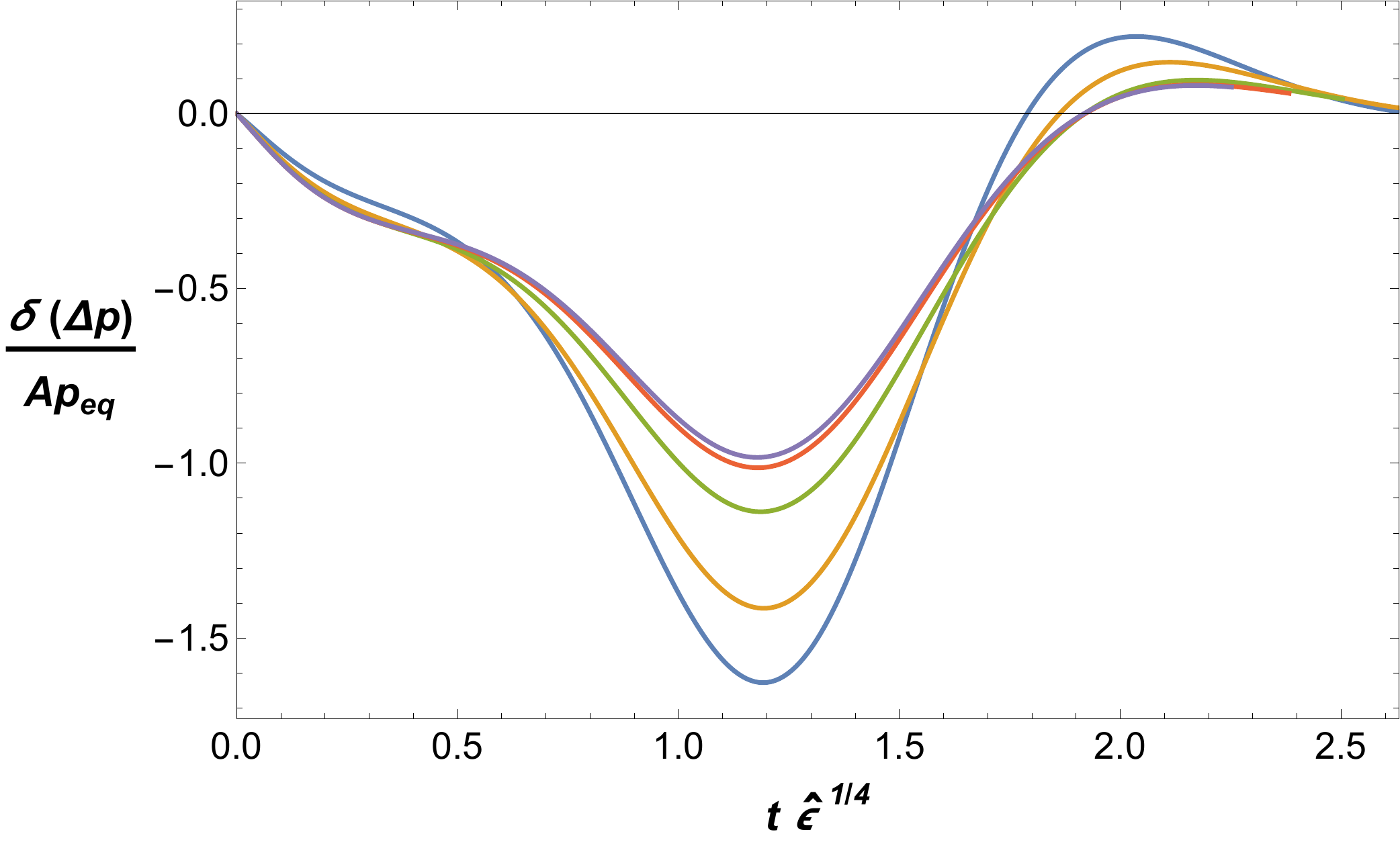}
\vskip 0.01\textwidth
\includegraphics[height=0.29\textwidth]{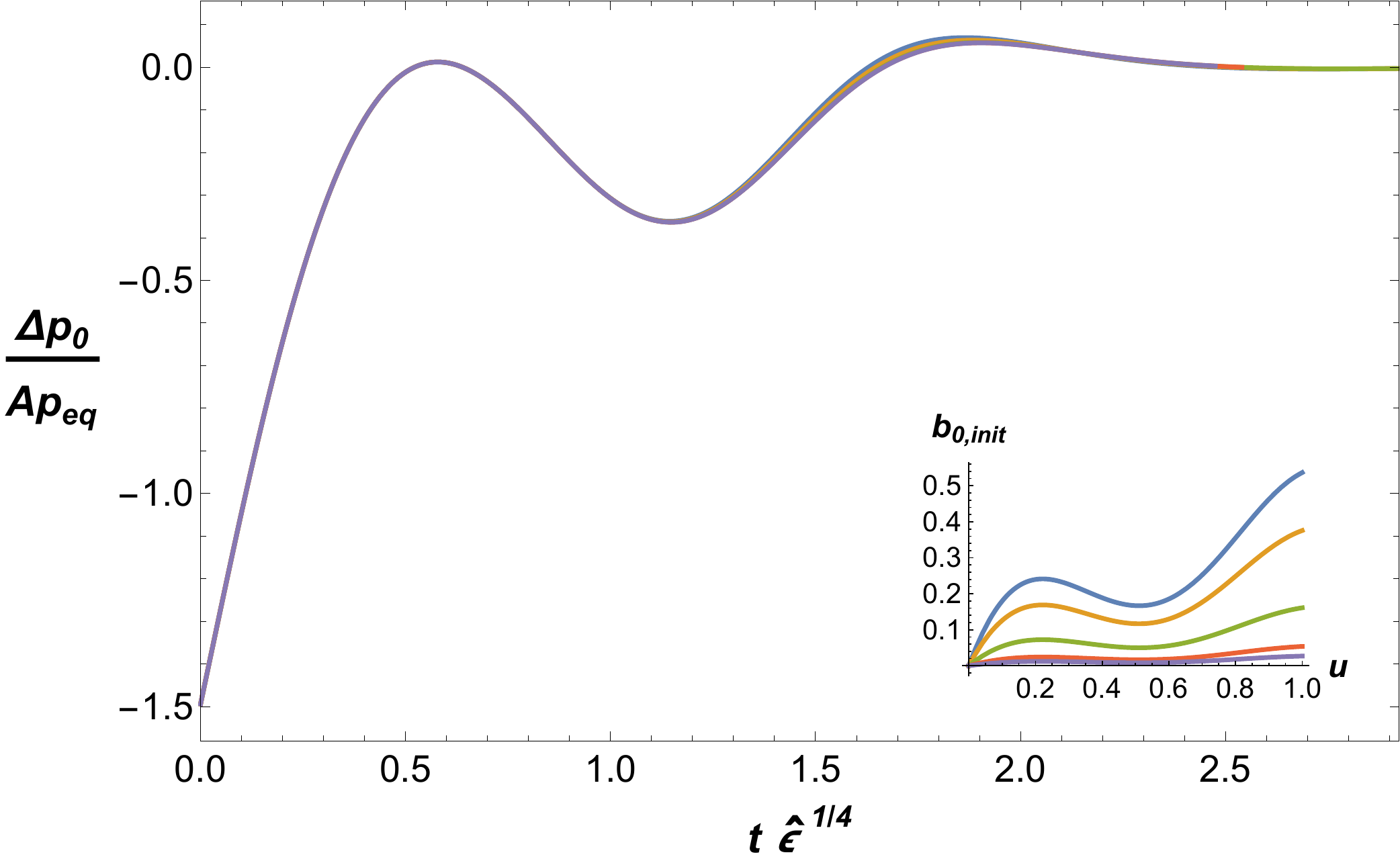}
\hskip 0.02\textwidth
\includegraphics[height=0.29\textwidth]{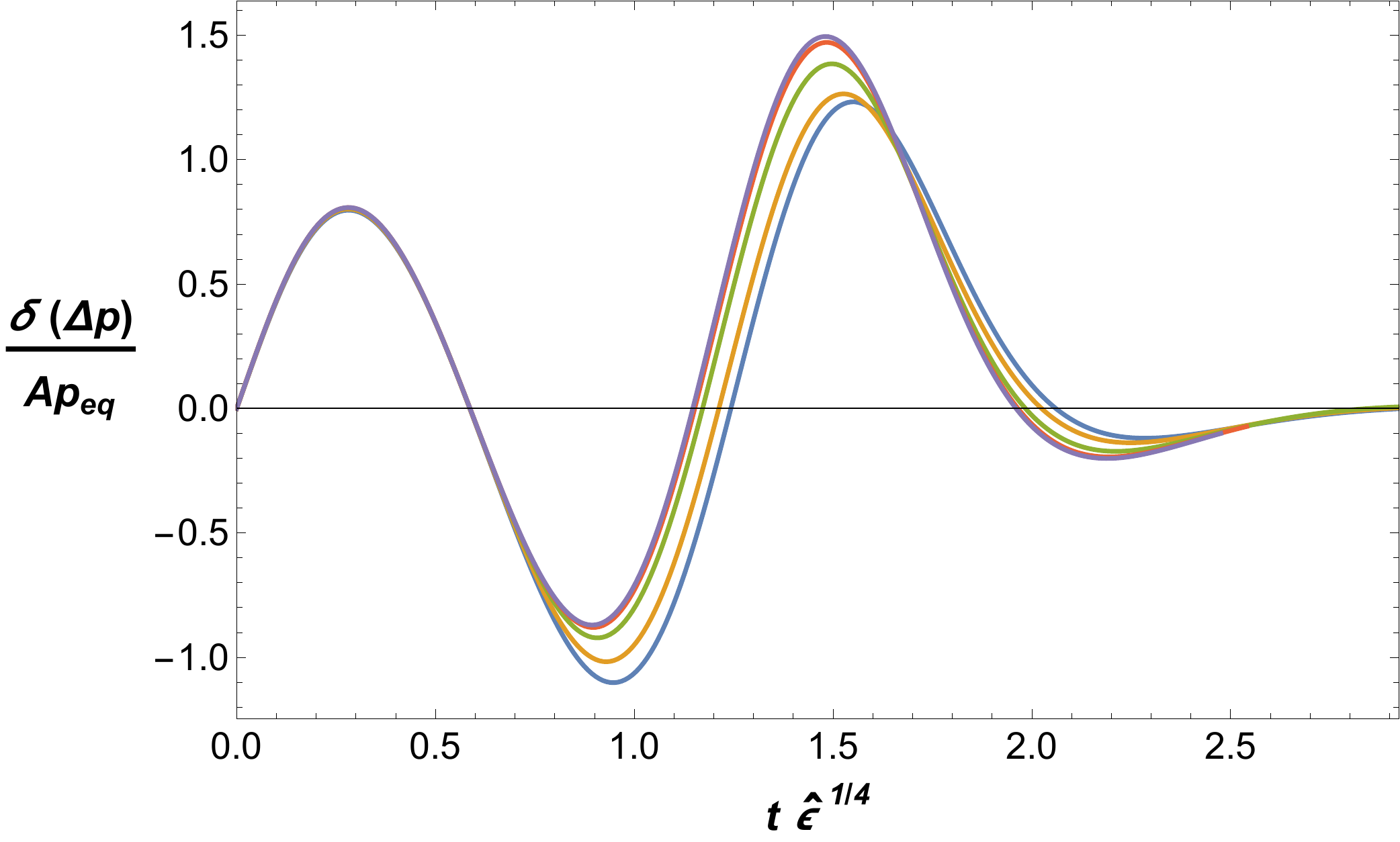}
\caption{Evolution of the pressure anisotropy for four different families of initial conditions. These are generated by specifying a functional form of the initial condition ${\cal F}(u)$ and multiplying this function by a scaling factor $\cA=1,\, 2\,, ...,\, 20$ as in \eno{scaling} (the different functions are shown in the insets in the left panels). In each row, the left panels shows the evolution of anisotropy for $\N=4$ SYM 
and the left one shows the evolution of the leading $\lGB$ correction for each of the families. All curves are rescaled by the value of ${\cal A}$. }
\label{p1}
\end{figure*}

The specification of the boundary condition \eqref{DeltaA4} and the initial condition \eqref{db IC} determines the linearised solutions
uniquely.
 Implementing the algorithm described above for a variety of choices of background initial conditions $b_{0, {\rm init}}$, 
we obtain the numerical results depicted in Fig. \ref{p1} (all our numerics are produced with $a_0^{(4)} = - 1/2$). 
The 4 rows in the figure correspond to 4 different families of initial conditions. Each of the families was generated by specifying a given functional form of the initial condition ${\cal F}(u)$ (as described in Sec. \ref{preliminaries}), and different members of a family correspond to choosing different amplitudes 
such that
\begin{equation}
\label{scaling}
 	b_{0, {\rm init}}(u) = \cA \,{\cal F}(u)\, ,
 \end{equation} 
\noindent for values of $\cA$ ranging from $1$ to $20$. 

In the left column of Fig. \ref{p1} we show the evolution for the background pressure anisotropy
 $\Delta p_0$, scaled by the amplitude $\cA$.  The functional forms of $b_{0, {\rm init}}(u)$ are shown in the insets in each of the panels. We can see that, in agreement with the results of \cite{HoloLinPRL,HoloLinJHEP}, this evolution is, to a remarkably good approximation, linear for the whole range of amplitudes, as evident from the fact that all the curves $\Delta p_0/ \cA$ lie approximately on top of each other. This is the case even for large anisotropies where one would be expect to have a non-linear evolution. 

In the right column of Fig. \ref{p1} we show the perturbed anisotropy $\delta \left(\Delta p\right)$ for the same family of solutions studied in the left panels. Unlike the $\lGB=0$ case, the time evolution does not scale with $\cA$ for all values of $\cA$. While for moderate values of $\cA \sim 2$, the perturbed anisotropy does exhibit a clear linear behaviour, the more extreme values of $\cA\sim20$ show  deviations from linearity. Therefore, while the isotropisation dynamics at $\lGB=0$ are effectively linear for the whole range of initial conditions we have explored, the higher derivative corrections only exhibit the approximately linear behaviour for large, but not arbitrarily large, anisotropies. Note that the apparent linear behaviour still occurs in a regime of anisotropies where full non-linear evolution would be expected, since the initial conditions that exhibit scaling still have rather large values of anisotropy $\delta \left(\Delta p \right)/p_{\rm eq}>1$. 
 
\begin{figure*}[ht!]
\centering
\includegraphics[height=0.29\textwidth]{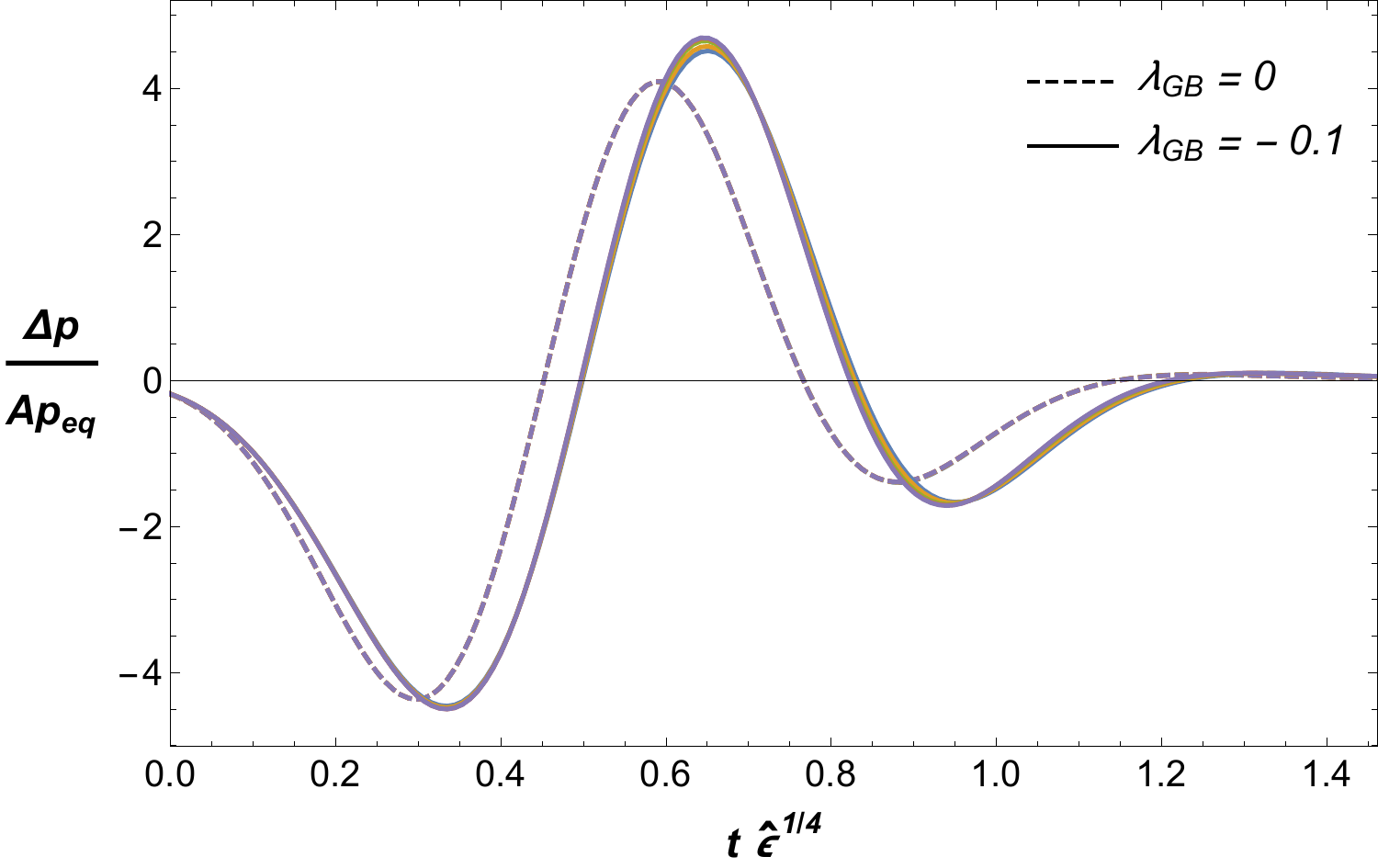}
\hskip 0.02\textwidth
\includegraphics[height=0.29\textwidth]{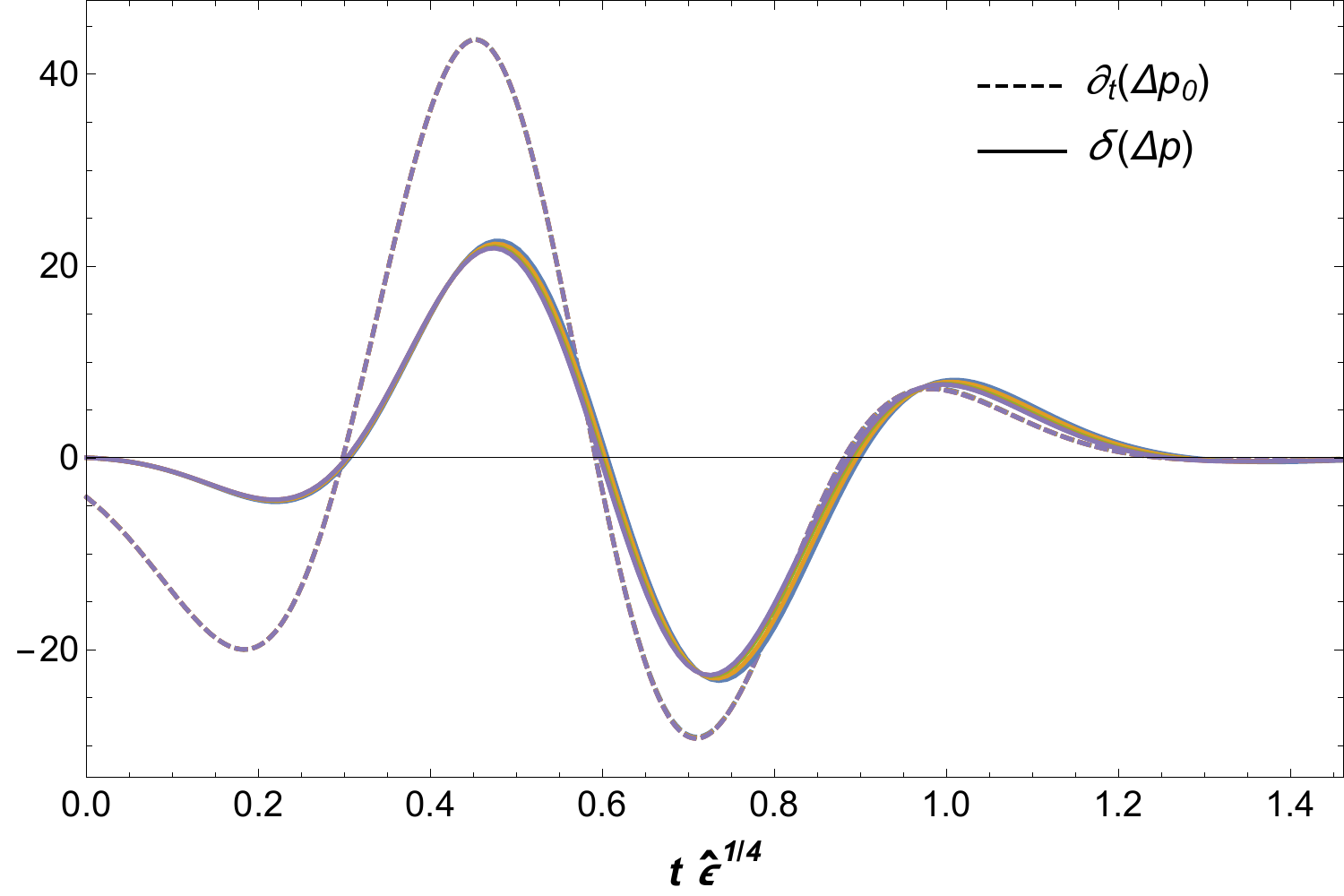}
\vskip 0.01\textwidth
\includegraphics[height=0.29\textwidth]{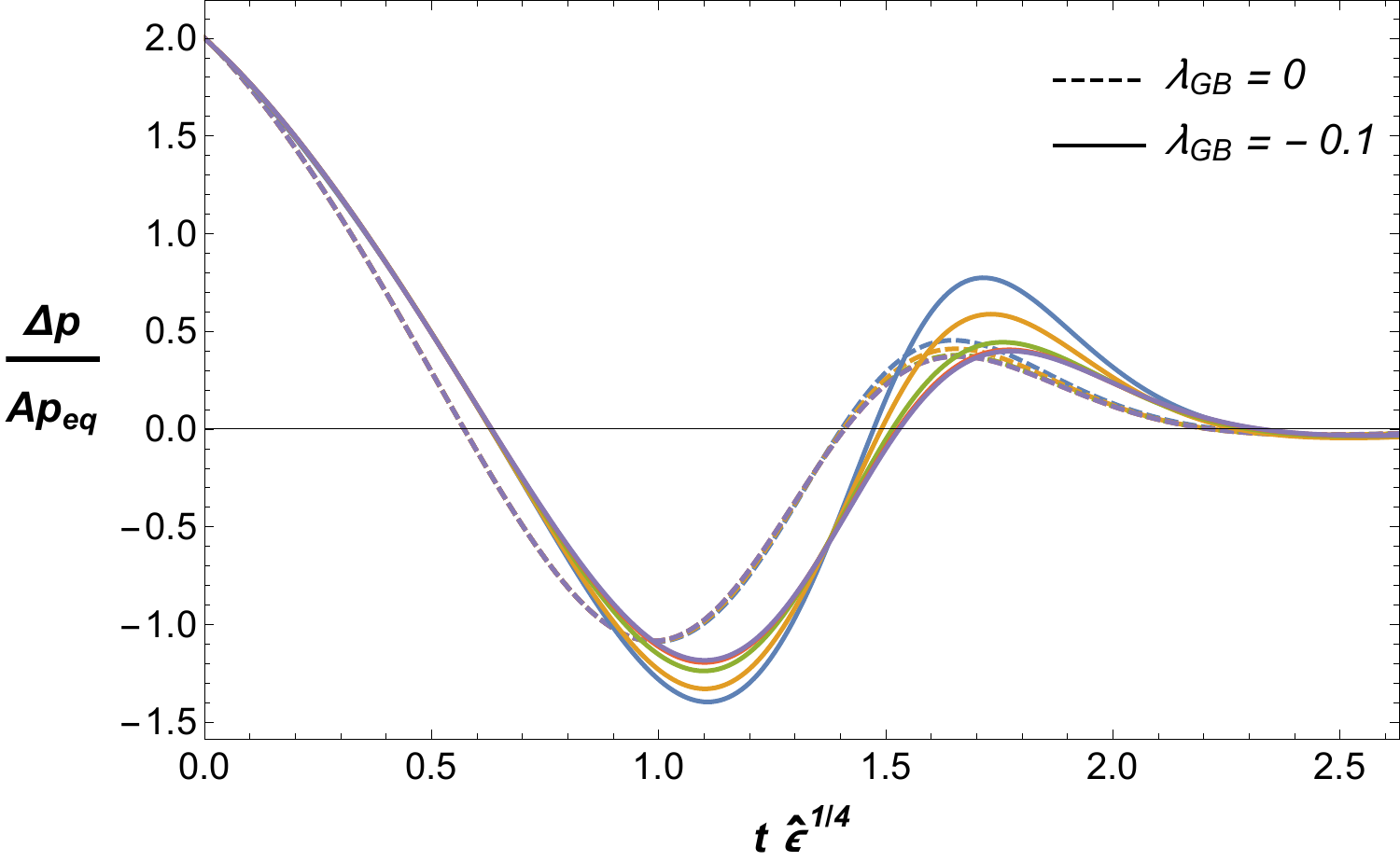}
\hskip 0.02\textwidth
\includegraphics[height=0.29\textwidth]{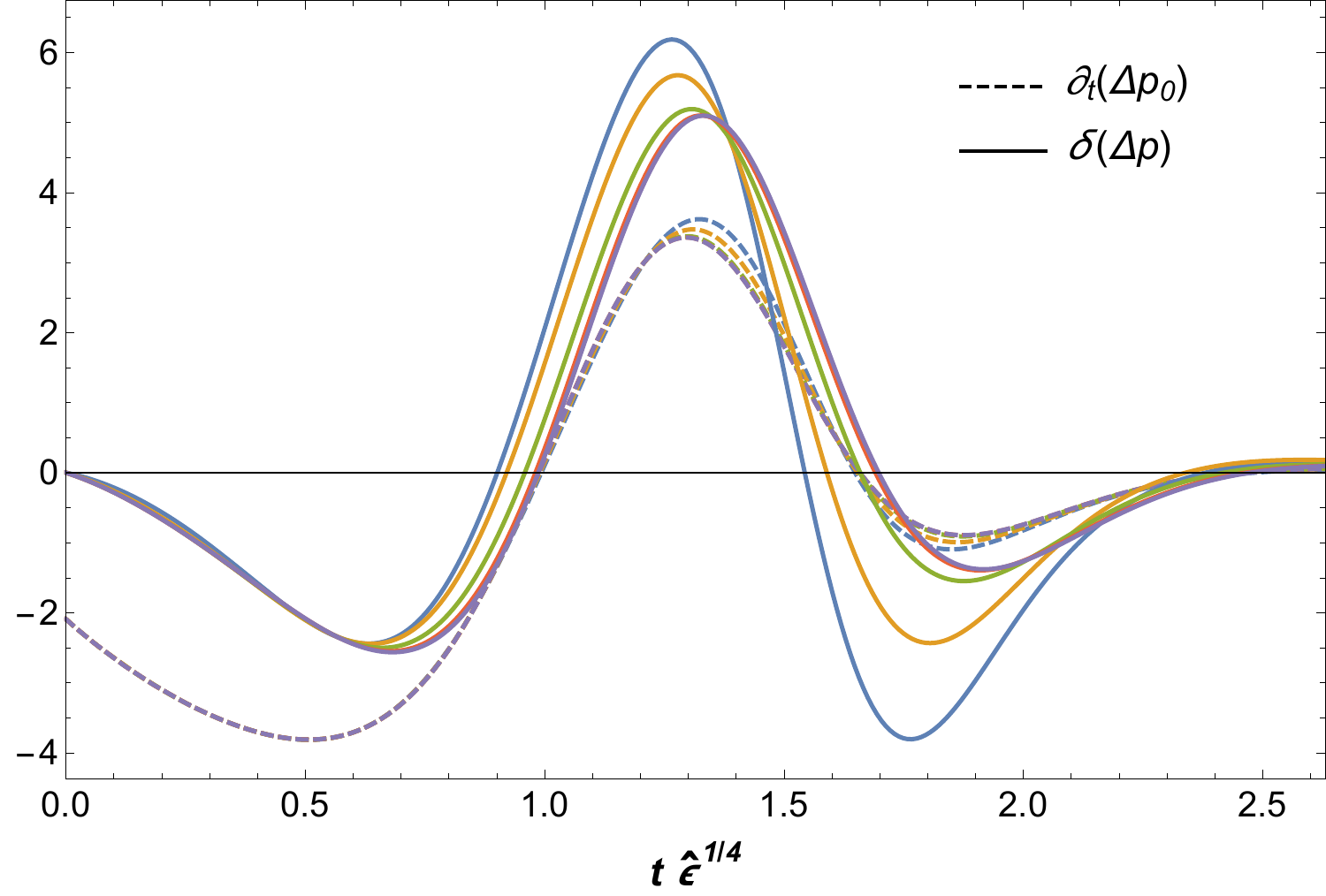}
\vskip 0.01\textwidth
\includegraphics[height=0.29\textwidth]{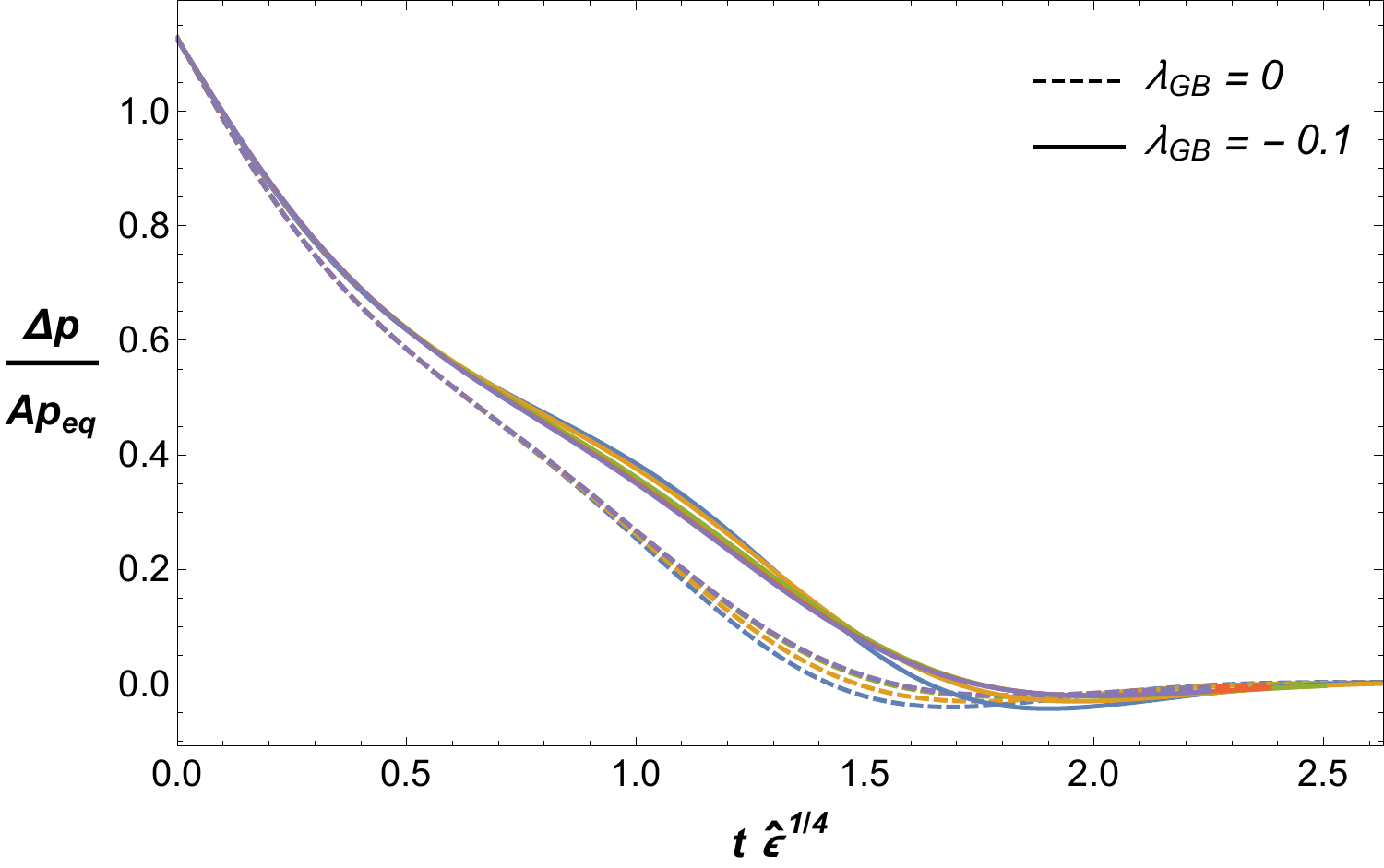}
\hskip 0.02\textwidth
\includegraphics[height=0.29\textwidth]{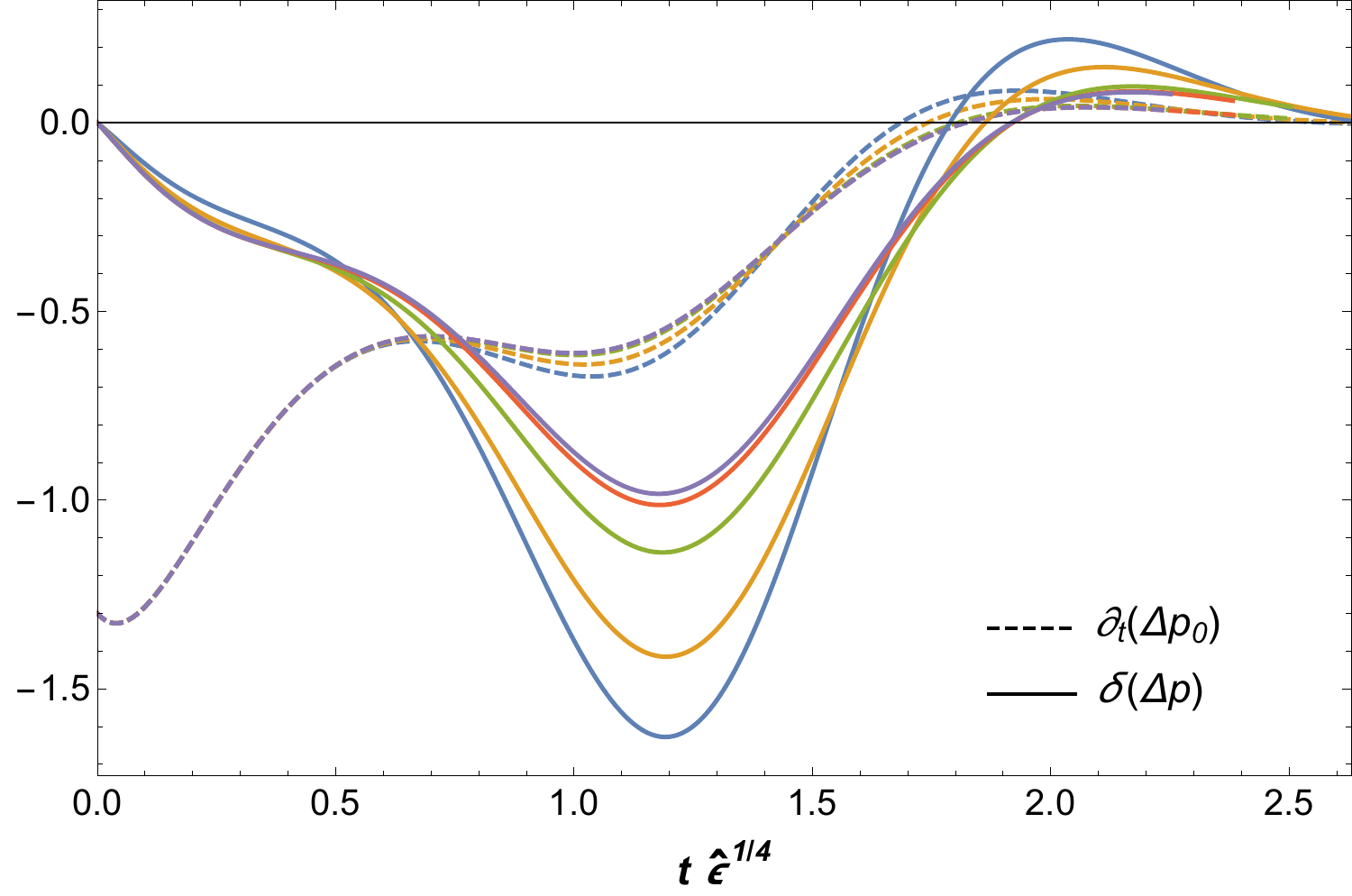}
\vskip 0.01\textwidth
\includegraphics[height=0.29\textwidth]{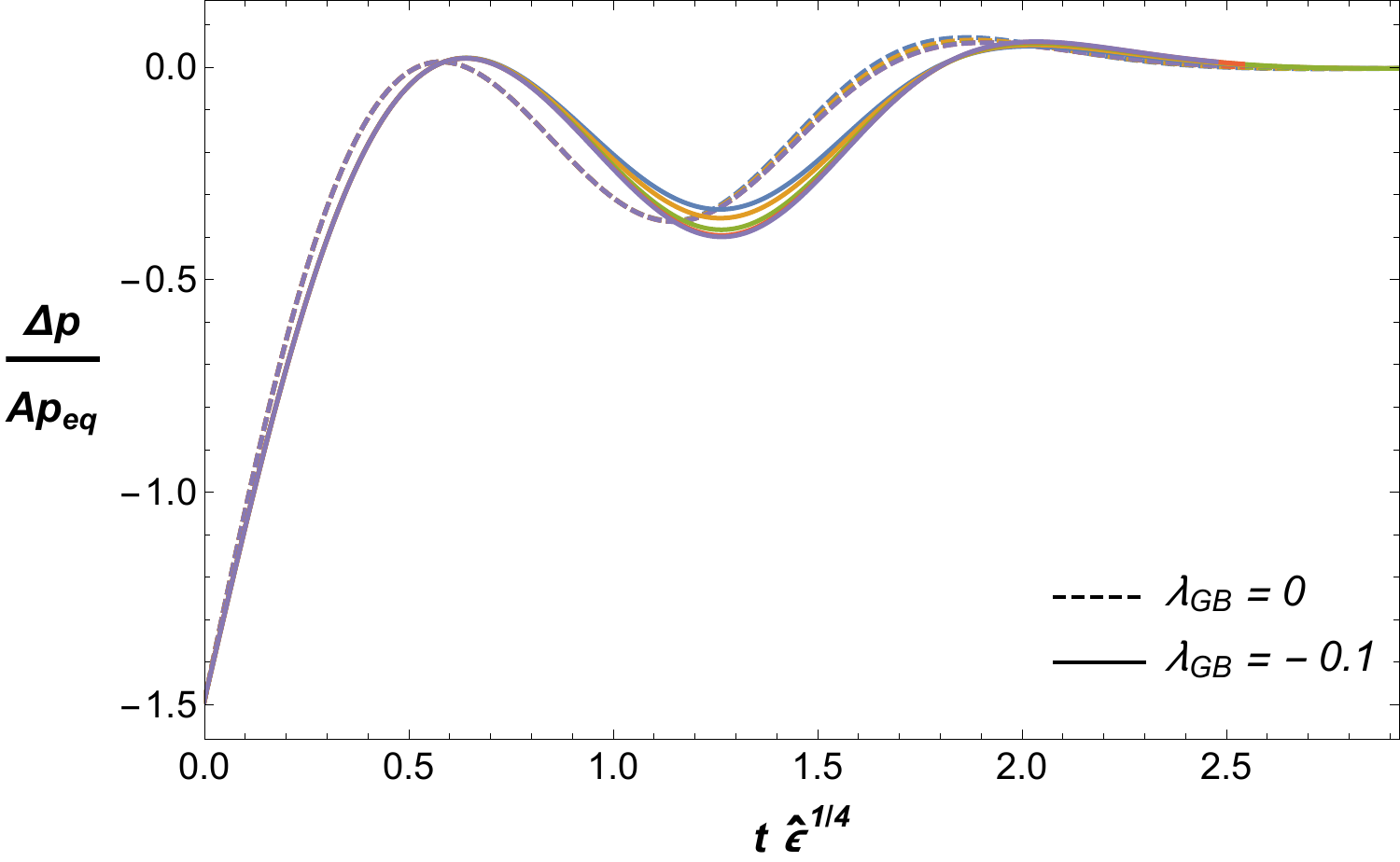}
\hskip 0.02\textwidth
\includegraphics[height=0.29\textwidth]{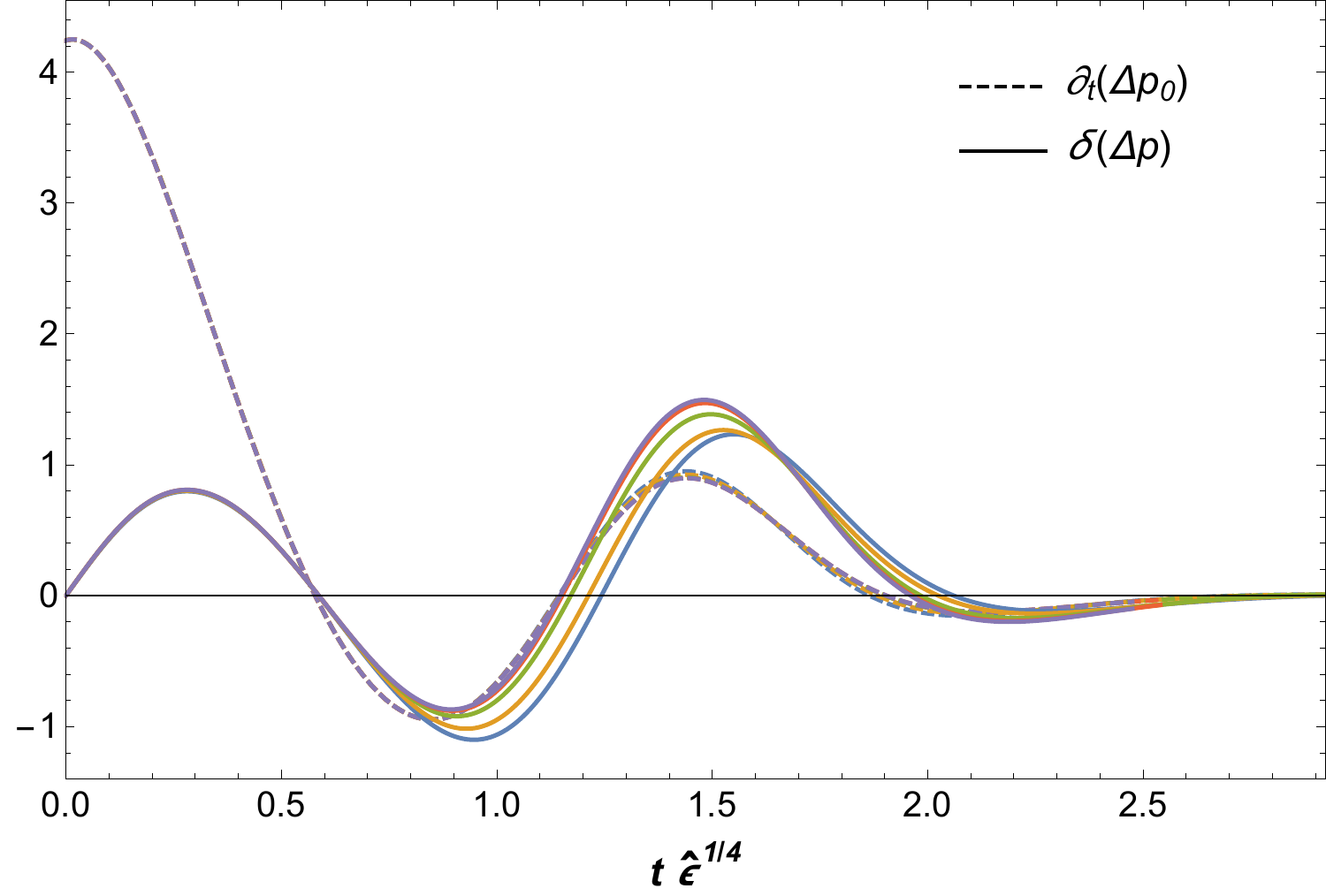}
\caption{Left: Evolution of the full pressure anisotropy for four different families of initial conditions for a fixed value of $\lGB=-0.1$ (solid curves), compared to the time evolution in $\mathcal{N}=4$ SYM (dashed curves).  Right: Comparison of the perturbed pressure anisotropy $ \delta (\Delta p) (t)$ and the time derivative of the background pressure anisotropy $\partial_t \Delta p_0 (t) $ for the same families of initial conditions. All curves are scaled with the multiplicative factor $\cA$ which relates the different members of a family (\eno{scaling}).}
\label{p2}
\end{figure*}

In Fig. \ref{p2} we show the direct effect of higher curvature corrections on the isotropisation process for the same family of initial conditions studied in Fig. \ref{p1}. In the left column, we show the full pressure anisotropy $\Delta p/p_{\rm eq}$ assuming a fixed value of $\lGB=-0.1$ (solid lines) in comparison with the $\N=4$ SYM case (dashed lines) for different families of initial conditions.  As in Fig. \ref{p1}, the anisotropy  is scaled by $\cA$ for each configuration. As we can see, for all families, the effect of the higher curvature corrections is to approximately shift the pressure anisotropy profile towards later times. We can quantify the approximate shift between $\Delta p$ and $\Delta p_0$ by comparing the time derivative of the background anisotropy, $\partial_t \Delta p_0 (t)$,  with the perturbed anisotropy $ \delta \left(\Delta p  \right)(t)$, which we do in the right column of Fig. \ref{p2}. To leading order in $\lGB$, the full evolution may be understood as a shift of the background anisotropy, $\Delta p(t) = \Delta p_0( t +  \lGB \Delta \tilde t(t)) $ with $\Delta \tilde t$ a slowly varying function of $t$, as long as 
\begin{equation}\label{sign test for shift}
{\rm sign \,} \left[\partial_t \Delta p_0 (t)\right] = {\rm sign \,} \left[\delta\left( \Delta p \right)(t)\right]\,.
\end{equation}
The inspection of Fig. \ref{p2} shows that \eno{sign test for shift} holds for almost the entire time evolution, at least as long as the anisotropy is not too large. Note that this statement is independent of value of $\lGB$ and hence it holds for all (small) values of $\lGB$. This observations leads us to conclude that for all initial conditions we have studied, the effect of the higher curvature corrections is to delay or advance the isotropisation, depending solely on the sign of $\lGB$. In the next Section we will show that for those initial conditions in which the full time evolution is approximately linear, this correlation between the signs of $\lGB$ and the shift always holds, irrespective of the initial conditions. 
 
%
%

\section{Analysis of the results with QNM expansions}
\label{sec:QNM}

The results from the previous section indicate that, at linear order in $\lambda_{GB}$, the evolution of the pressure 
anisotropy in Gauss-Bonnet gravity behaves in an approximately linear fashion, similarly to the case of $\lambda_{GB} = 0$ described in \cite{HoloLinPRL,HoloLinJHEP}, although for a smaller range of initial pressure anisotropies. 
This suggests that the QNM of the final state can capture the full time evolution well, even at early times. 
In this section we check this hypothesis and conclude that, indeed, for amplitudes for which the linear behaviour in $\delta \left(\Delta p\right)$
is observed, its time evolution is well-described by the QNM's of the Gauss-Bonnet black hole.
Therefore, in the regime in which the QNM description is applicable, the dynamics of the system are 
largely simplified and, to a good approximation, the results can be obtained without resorting to full numerical time evolution. 
Armed with this simplification, we argue that the introduction of the Gauss-Bonnet term induces a 
time-shift of $\Delta p$ with a definite sign, consistent with our numerical findings in Sec. \ref{sec:time evolution GB}. Our argument applies to configurations which satisfy initial conditions more general than \eno{db IC}, suggesting that the observed time shift in the pressure anisotropy is a generic feature of finite coupling corrections.

\subsection{QNM in AdS/CFT}

Generically, QNM's are linearised fluctuations around a black hole background which satisfy ingoing boundary conditions at the horizon\footnote{Note that the ingoing boundary condition translates simply into regularity in coordinates of the form \eno{BulkAnsatz}.} and an appropriate boundary condition in the UV. In the context of AdS/CFT, the interpretation of the slow/fast fall-offs of the fields at infinity as sources/vevs for the dual operators indicates what this boundary condition should be if one is to interpret the QNM as poles of the corresponding 2-point function: we simply set to zero the coefficient which corresponds to the source. The frequency of these excitations is generally complex, and their imaginary part naturally provides a time scale for the decay of the perturbation. The longest time-scale is then controlled by the lowest lying QNM's and these govern the late-time dynamics. See \cite{Berti:2009kk} for a review. 

In the context of holographic isotropisation, the relevant QNM are the linearised fluctuations of the field $B$ which quantifies the anisotropy of the brane. It is easy to check that at the linear level these fluctuations decouple and can be studied on their own. In Fourier space, the perturbation equation is an ODE which defines an eigenvalue problem for the (complex) frequency $\omega$. The frequency spectrum is fixed once the boundary conditions at the horizon and the boundary are provided. As mentioned above, the boundary condition at the horizon is simply regularity there, which translates into $B$ having a regular power series expansion in $(r-r_H)$. On the boundary, we demand the induced metric to be fixed $B = 0$, since this gives the QNM the interpretation of poles of the stress-tensor correlator.  

The QNM's relevant for isotropisation have been studied in the case of AdS-Schwarzschild planar black hole in \cite{Kovtun:2005ev}, and more recently, their Gauss-Bonnet counterparts have been obtained at finite $\lambda_{GB}$ \cite{AndreiGB}. We refer the reader to these references for details of their calculation and the specific values of their frequencies.

\subsection{Matching of QNM in AdS}

Our approach is based on the work of \cite{HoloLinPRL,HoloLinJHEP}, where it was shown that the time evolution of the pressure anisotropy in $\mathcal{N}=4$ SYM, even in highly anisotropic cases, can be well captured by a truncated expansion in QNMs, i.e. by linearising Einstein equations around the final equilibrium state of the AdS-Schwarzschild black hole. 
The idea is to decompose the solution for $b_0(t,u)$ in QNM's as follows:
\eqn{QNMExpZeroth}
{b_0(t,u)={\rm Re}\left[\sum\limits_{i=1}^NC_i^{(0)}\phi_i^{(0)}(u)\,e^{ -i \omega_i^{(0)}  t}\right]\,,}
\noindent where $N_{\rm QNM}$ is the number of QNMs we will use in practice, $C_i^{(0)}$ are (complex) 
coefficients, $\phi_i^{(0)}(u)$ are the wave functions of the $i-$th QNM of AdS-Schwarzschild in the scalar channel at zero momentum\footnote{Note that in the zero spatial momentum all channels of excitations of the stress tensor become identical.} (normalised so that at the horizon $\phi_i^{(0)}(1)=1$), and $\omega_i^{(0)}$ are the corresponding complex frequencies. We can find the coefficients $C_i^{(0)}$ by requiring \eno{QNMExpZeroth} match our initial conditions at $t=0$ as closely as possible -- in practice, this can be done by performing a multilinear regression on some grid in the interval $u\in[0,1]$.
After such procedure, 
\eno{QNMExpZeroth} provides a very  good approximation for the full time evolution already for a rather small number of QNM, and for fairly large values of anisotropies, as discovered in \cite{HoloLinPRL,HoloLinJHEP}.

\subsection{Matching of QNM in Gauss-Bonnet}

Following the ideas from the previous section, we would like to decompose the time evolution of $b(t,u)$ in QNMs in the Gauss-Bonnet case as well:
\eqn{QNMExpGB}
{b(t,u)={\rm Re}\left[\sum\limits_{i=1}^NC_i\phi_i(u)\,e^{ -i \omega_i  t}\right]\,,}
where now the lack of $(0)$ subscripts indicates that these quantities are the full expressions in the Gauss-Bonnet case. We will expand those linearly in $\lambda_{GB}$:
\eqn{QNMExpModes}
{\begin{aligned}
\phi_i & = \phi_i^{(0)}+\lambda_{GB}\,\delta \phi_i \,,\cr
C_i & = C_i^{(0)}+\lambda_{GB}\,\delta C_i \,,\cr
\omega_i & = \omega_i^{(0)}+\lambda_{GB}\,\delta \omega_i \,.
\end{aligned}}
Plugging \eno{QNMExpModes} in \eno{QNMExpGB}, expanding in $\lambda_{GB}$, and remembering the definition of $\delta b$, we have:
\eqn{DeltabExp}
{\delta b(t,u)={\rm Re}\left[\sum\limits_{i=1}^N e^{ -i \omega_i^{(0)}  t} \left( \delta C_i\phi_i^{(0)} +\delta\phi_i C_i^{(0)} -it\delta \omega_i C_i^{(0)} \phi_i^{(0)}\right) \right]\,.}
Here we note that, even though we have a term linear in $t$, this expansion is still valid up to arbitrarily large times, since $\delta b(t,u)$ is multiplied by an infinitesimally small $\lambda_{GB}$.

In order to compute $\delta \omega_i$ and $\delta \phi_i(u)$, we have numerically solved the Gauss-Bonnet QNM equations 
at finite $\lambda_{GB}$\footnote{We performed the numerics 
by discretising the eigenvalue problem using pseudospectral methods with a Chebyshev lattice. 
This allows us to go up to 
$N_{\rm QNM}=10$.} and evaluated the derivatives of $\omega_i$ and $\phi_i$ with respect to $\lambda_{GB}$, at $\lambda_{GB}=0$.
Care must be taken when computing $\delta \phi_i$ and $\delta \omega_i$, as the former is a function of a dimensionful quantity, and the latter is a dimensionful quantity itself. Since our theories are finite temperature CFT's, these quantities are typically computed as dimensionless numbers, using the temperature to make them such. However, because of our requirement that the two systems, at zero- and nonzero-$\lambda_{GB}$, have the same energy densities (\eno{DeltaA4}), their final temperatures are not the same. Equating \eno{ET0} and \eno{ETlambda} we find the relation between these two temperatures to be 
\eqn{TGB}
{\begin{aligned}
T&=\frac{\left( 1+\sqrt{1-4\lambda_{GB}} \right)^{3/8}}{\pi\, 2^{3/8}} = T_0 \left(1 - \frac{3}{8} \lambda_{GB}\right) + \mathcal{O}\left( \lambda_{GB}^2 \right)\,.\cr
\end{aligned}} 
We will use this relation to relate the perturbations of the quasinormal modes at fixed energy density with those at fixed temperature. 

We start with the explicit expression for $\delta \omega_i$.
The numerically generated dimensionless frequency $\tilde \omega_i= \omega_i / (\pi \,T)$ as a function of $\lambda_{GB}$, can be expanded around $\lambda_{GB} = 0$, yielding:
\eqn{OmegaExp}
{\frac{\omega_i}{\pi \,T} = \tilde \omega_i^{(0)} + \lambda_{GB}\delta \tilde\omega_i\,.}
After multiplying this expression by $\pi T$, with $T$ given by  \eno{TGB}, we obtain
\eqn{OmegaExp2}
{\delta \omega_i =\pi T_0\left( \delta \tilde \omega_i-\frac{3}{8}  \tilde \omega_i^{(0)} \right)\,.}
Note that with the choice of $a_0$ in our numeric simulations, $T_0=1/\pi$. 
For  $\delta \phi_i$, the procedure is analogous. From the numerical computation of the QNM, we extract the variation of the wave function evaluated at the same argument
\eqn{PhiExp2}
{\phi_i(\tilde u) =  \phi^{(0)}_i(\tilde u)+\lambda_{GB} \delta \tilde \phi_i (\tilde u)\,.}
where $\tilde u = \pi T u$. Using \eno{TGB} and expanding to leading order we obtain 
\eqn{PhiExp3}
{\delta \phi_i(u) = \delta \tilde \phi_i (u) - \frac{3}{8} u \partial_u\phi_i^{(0)}(u)\,, }
where we have explicitly used that in our choice of units $T_0=1/\pi$.

To determine $\delta C_i$ we impose that at $t=0$  \eno{DeltabExp} matches the initial condition  \eno{db IC} for $\delta b$ as closely as possible.
As in the zeroth order case, we use multilinear regression to determine the best-fit coefficients. It is worth noting that even if we start with an initial condition  $b_{0 , {\rm init}}$ such that only certain zeroth order QNM is excited (and so the zeroth order time evolution of the pressure anisotropy is governed by that mode only), in the time evolution of the perturbations, more than one Gauss-Bonnet QNM will be excited. 

Having fixed all the components of \eno{DeltabExp} for $\delta b$ and using the QNM decomposition  \eno{QNMExpZeroth} for $b_0$, we use 
 \eno{DeltaDeltaP} for the change in the pressure anisotropy to arrive at 
\eqn{DeltaDeltaPExp}
{\delta(\Delta \hat p) = -3\,{\rm Re}\left[\sum\limits_{i=1}^N e^{ -i \omega_i^{(0)}  t}\left( \left(\frac{1}{2} -it\delta\omega_i \right)C_i^{(0)}\partial_u\phi_i^{(0)} + \delta C_i \partial_u\phi_i^{(0)}+ C_i^{(0)}\partial_u\delta\phi_i    \right) \right]\,,}
where the $u-$derivatives are evaluated at $u=0$. Similarly,  and for later reference, the background pressure anisotropy is given by 
\eqn{DeltaPExp}
{
{\Delta \hat p_0 = -3\,{\rm Re}\left[\sum\limits_{i=1}^N C_i^{(0)}\left(\partial_u\phi_i^{(0)}\right) e^{ -i \omega_i^{(0)}  t}
 \right]\,.}
}
%

\begin{figure*}[t]
\centering
\includegraphics[height=0.29\textwidth]{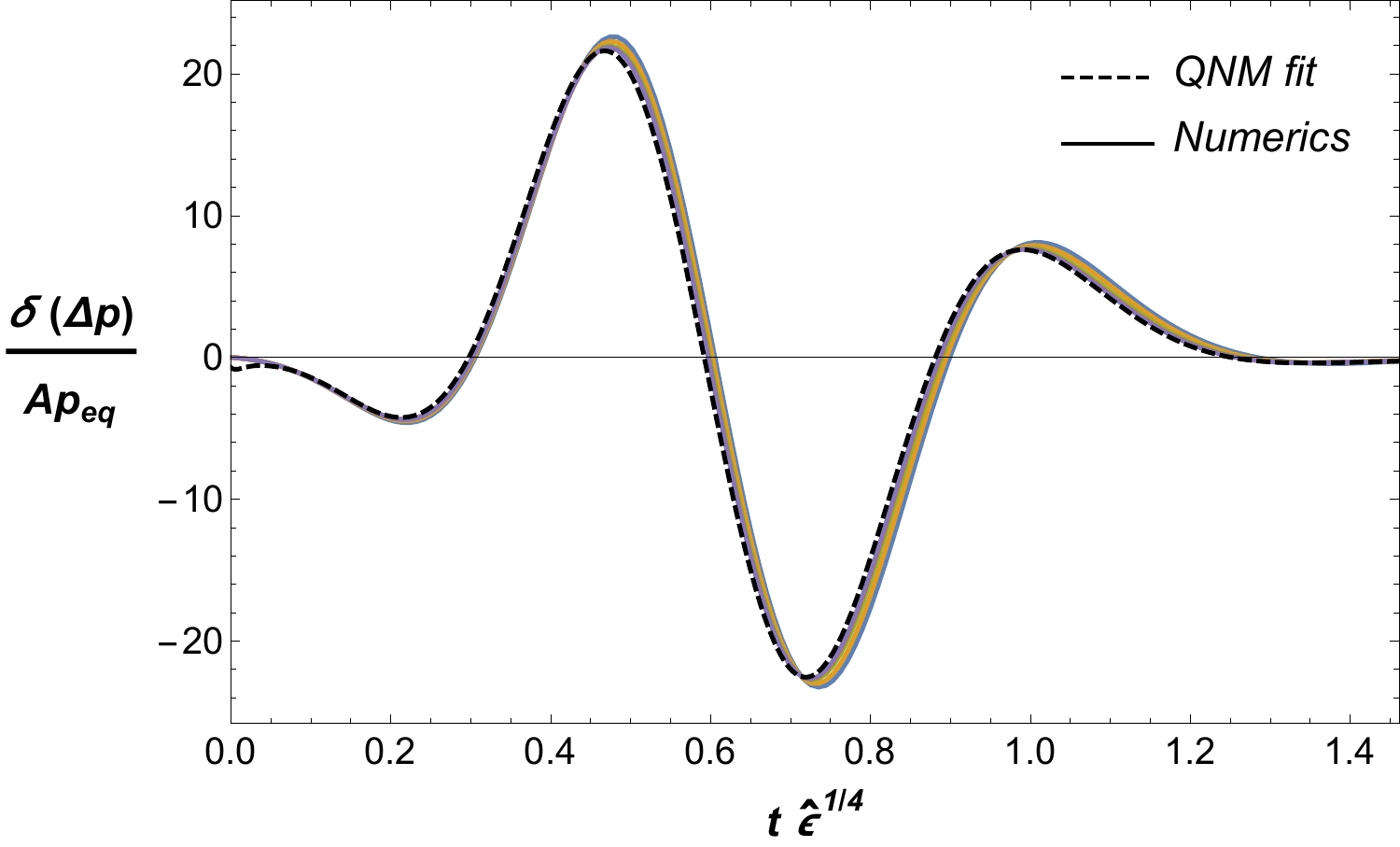}
\hskip 0.02\textwidth
\includegraphics[height=0.29\textwidth]{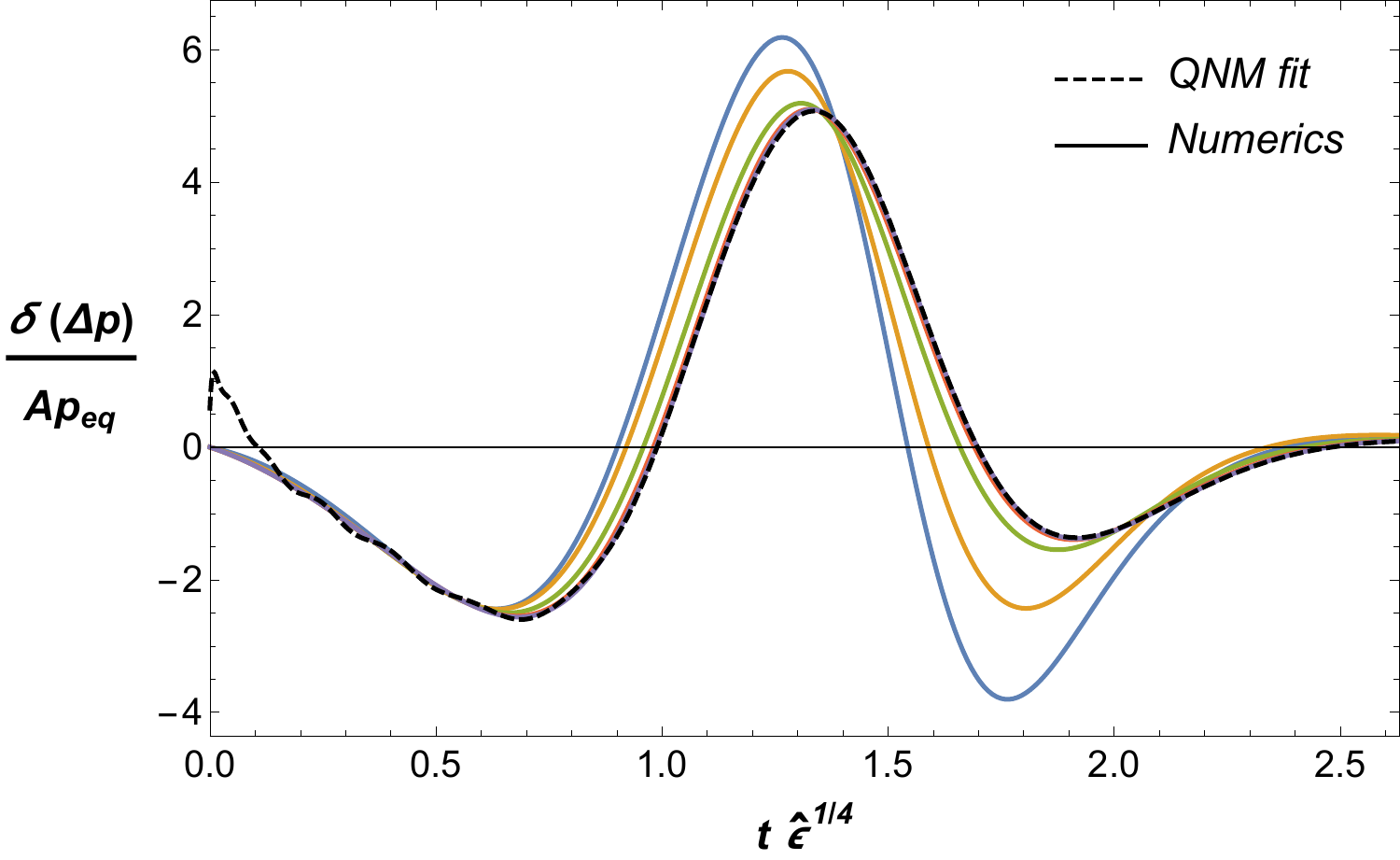}
\vskip 0.01\textwidth
\includegraphics[height=0.29\textwidth]{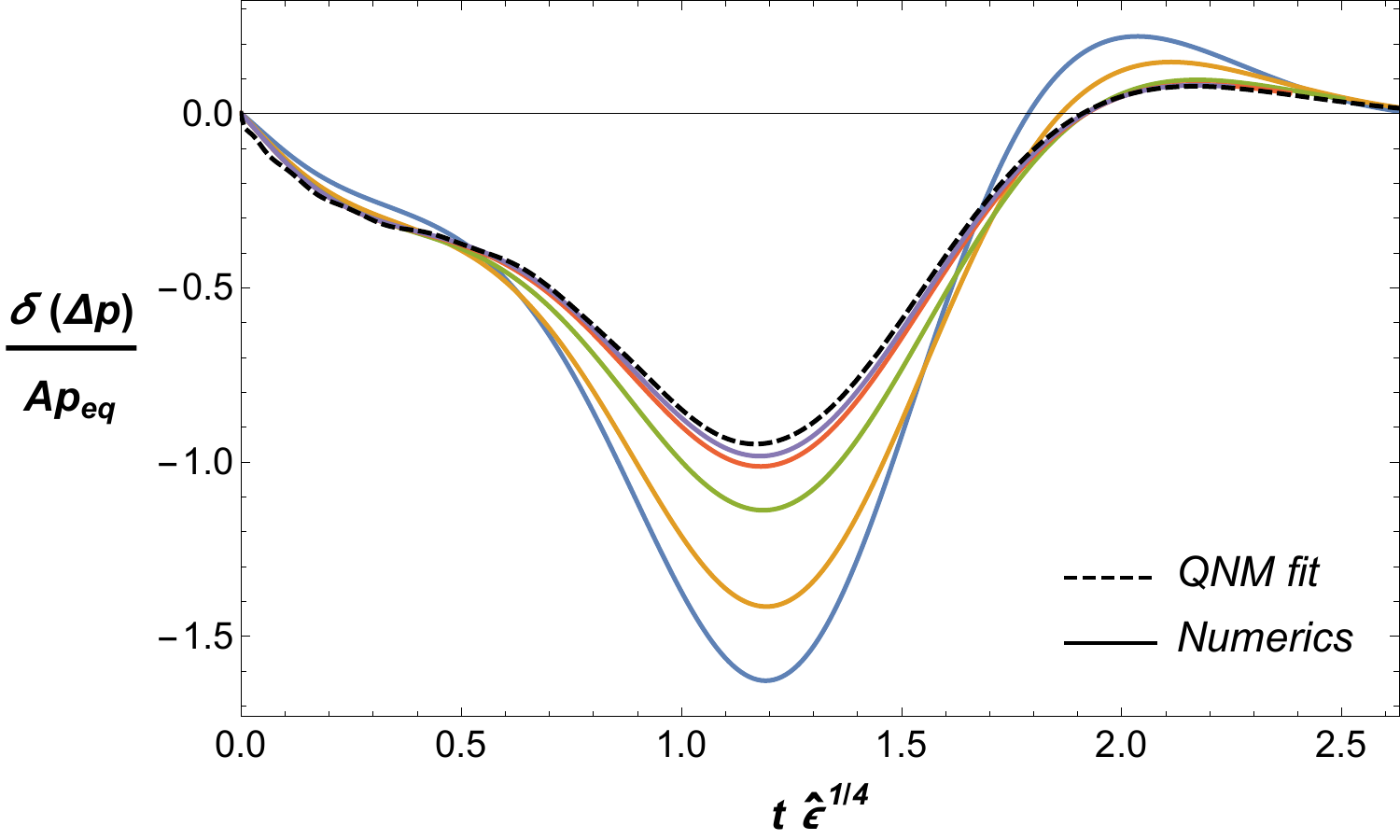}
\hskip 0.02\textwidth
\includegraphics[height=0.29\textwidth]{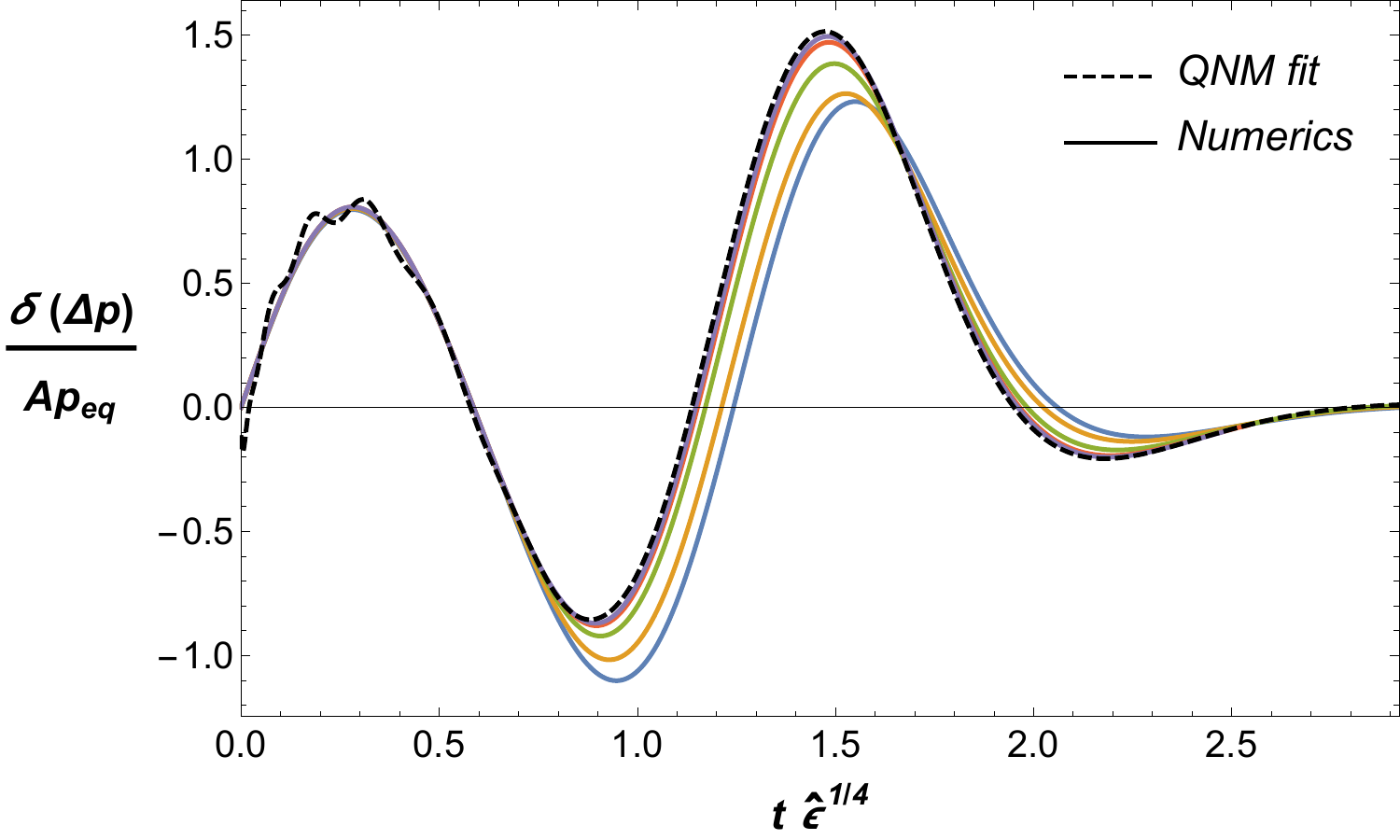}
\caption{ Perturbed pressure anisotropy $\delta \left( \Delta p\right)$ given by the QNM expansion \eqref{DeltaDeltaPExp} (dashed curves), compared to the full numerical results (solid lines) for four different families of initial conditions related by a multiplicative factor (\eno{scaling}). The quasinormal mode evolution describes the numerical results well for those amplitudes that exhibit effective linear behaviour, as shown in Fig. \ref{p1}.  }
\label{QMNfit}
\end{figure*}

In Fig.~\ref{QMNfit} we compare the time dependence of $\delta \left( \Delta p \right)$ as predicted by the QNM fit, \eno{DeltaDeltaPExp}, with our numerical simulations for the four families of solutions studied in Sec. \ref{sec:time evolution GB}. We see that, as long as the initial conditions are such that the system behaves effectively linearly, \eno{DeltaDeltaPExp} provides a good description of the full time evolution. However, as the anisotropy increases to very large values, clear deviations from \eno{DeltaDeltaPExp} are observed, consistent with the absence of linear behaviour observed in Fig. \ref{p1}.
%

\subsection{Time shift from QNM}
\label{TimeShift}

We will now argue that the introduction of the Gauss-Bonnet term induces a time shift in the pressure anisotropy
in the sense of \eno{sign test for shift}.
%
Our derivation will be valid for those initial conditions in which the full evolution can be described via QNM, which, as indicated in
 the previous section, can be valid even for large initial pressure anisotropies, 
  $\delta \Delta p/p_{\rm eq} \sim \mathcal{O} (1)$. 
 We will come back to the validity of the QNM approximation 
 in  Sec.~\ref{sec:discussion}.

As we have already mentioned,
 the expansion of the initial conditions in terms of a (finite) set of $N$ modes is performed via a multilinear regression 
on a given grid $\{u_m\}$, $m=1,...,M$. In this section we will provide explicit expressions for this procedure and use them to determine a relation between 
the coefficients $\delta C_i$ and $C_i^{(0)}$, introduced in \eno{QNMExpModes}, which control the QNM approximation of the anisotropic metric coefficient $b$,
as shown in \eno{QNMExpGB}.  This relation will allow us to express the change of the pressure anisotropy $\delta \left( \Delta p\right)$ directly in terms of initial conditions,
which will be the basis to argue that the relation \eqref{sign test for shift} is true.
Althought it is possible to find an equivalent formulation in terms of a basis of functions in the $u$-interval, 
our analysis in the discretised grid is closer to the actual numerical procedure we employ to approximate the time evolution via QNM, and, for this reason, we will focus on that approach.

For convenience, let us first introduce some notation:
\eqn{BookKeeping}
{\begin{aligned}
&\mathcal{C}_1 = {\rm Re} \,C_1^{(0)},\dotsc, \,\mathcal{C}_N= {\rm Re} \,C_N^{(0)}\,,\cr
&\mathcal{C}_{N+1} = {\rm Im}\, C_1^{(0)},\dotsc,\,\mathcal{C}_{2N}= {\rm Im} \,C_N^{(0)}\,,\cr
&\delta\mathcal{C}_1 = {\rm Re}\,\delta C_1,\dotsc, \,\delta\mathcal{C}_N= {\rm Re}\,\delta C_N\,,\cr
&\delta\mathcal{C}_{N+1} = {\rm Im}\,\delta C_1,\dotsc,\,\delta\mathcal{C}_{2N}= {\rm Im}\,\delta C_N\,,\cr
&x_1^{(m)}={\rm Re}\,\phi_1^{(0)}(u_m),\dotsc,\,x_N^{(m)}={\rm Re}\,\phi_N^{(0)}(u_m)\,,\cr
&x_{N+1}^{(m)}=-{\rm Im}\,\phi_1^{(0)}(u_m),\dotsc,\,x_{2N}^{(m)}=-{\rm Im}\,\phi_N^{(0)}(u_m)\,,\cr
&\delta x_1^{(m)}={\rm Re}\,\delta\phi_1(u_m),\dotsc,\,\delta x_N^{(m)}={\rm Re}\,\delta\phi_N(u_m)\,,\cr
&\delta x_{N+1}^{(m)}=-{\rm Im}\,\delta\phi_1(u_m),\dotsc,\,\delta x_{2N}^{(m)}=-{\rm Im}\,\delta\phi_N(u_m)\,,\cr
&y^{(m)}(\mathcal{C}_1\,,\dotsc\,,\mathcal{C}_{2N})=\delta b_{\rm init} (u_m)-\sum\limits_{n=1}^N\mathcal{C}_n \delta x_n^{(m)} \,,
\end{aligned}}
where $\delta b_{\rm init} (u)$ is a particular initial condition for $\delta b$ at $t=0$. Note that the quantities $x_{n}^{(m)}$ depend only on QNM's of AdS-Schwarzschild and are, therefore, independent of initial conditions. For notational convenience we package $\{\mathcal{C}_n\}$, $\{\delta \mathcal{C}_n\}$ and $\{x_{n}^{(m)} \}$ into $2N$-vectors  $\mathcal{C}$, $\delta \mathcal{C}$ and $x^{(m)}$, respectively, and $\{y^{(m)}\}$ into an $M$-vector $y$. 

With this notation, our multilinear regression problem of approximating $b_{0,{\rm init}}(u)$ with \eqref{QNMExpZeroth} at $t=0$ and $\delta b_{\rm init}(u)$ with \eqref{DeltabExp} at $t=0$ is equivalent to determining the coefficients $\mathcal{C}_{n}$ and $\delta \mathcal{C}_{n}$ by minimising the square errors on our grid:
\eqn{DeltabExplAlt}
{\begin{aligned}
{\underset{\{ \mathcal{C}_n\}}{\rm min}\sum\limits_{m=1}^M\left( \mathcal{C}\cdot x^{(m)}-b_0^{(m)}\right)^2\,,}
\cr
{\underset{\{\delta \mathcal{C}_n\}}{\rm min}\sum\limits_{m=1}^M\left(\delta \mathcal{C}\cdot x^{(m)}-y^{(m)}\right)^2\,,}
\end{aligned}}
where we have introduced $b_0^{(m)}=b_{0, {\rm init}}(u_m)$. The analytical solution to this problem is provided by the normal equation, which gives the following maximum likelihood estimate of the coefficients:
\eqn{SolRegression}
{\begin{aligned}
\mathcal{C} &= \rho \, b_0\, , \cr
\delta \mathcal{ C} &= \rho \, y \,,
\end{aligned}
}
where we packaged $b_0^{(m)}$ into an $M-$vector $b$, and where the $(2N \times M)$-dimensional matrix $\rho$ is given by
\eqn{NormalEq}
{\rho = \left( X^T \cdot X \right)^{-1}X^T \,,}
where we have rewritten $x_n^{(m)}$ as a matrix  $X_{mn}$, with the first index being the row index and the second one being the column one. The inverse in \eno{NormalEq} is meant to be the Moore-Penrose pseudoinverse in case $X^T X$ is non-invertible. 

The first equation in \eqref{SolRegression} provides an explicit expression for the coefficients $C_i^{(0)}$ in terms of the background metric function $b_0(0,u)$; therefore the vector $\mathcal{C}$ is fully specified by the initial conditions. We will now use the second equation in \eqref{SolRegression} to express $\delta \mathcal{C}$ as a function of the initial conditions as well. Examining the definition of $y$ in \eno{BookKeeping} we identify two distinct terms: the first term depends solely on the initial condition for perturbations $\delta b_{\rm init} (u)$, while the second term depends  on  the initial conditions for the background anisotropy (through $\mathcal{C}$) and on the modifications of the QNM wave functions, which are a property of the asymptotically late stable state and are common to all perturbations. Therefore, after packaging $\{\delta b_{\rm init}(u_m)\}$ into an $M$-vector $\delta b$, we reorganise the second equation in \eqref{SolRegression} as
\eqn{deltaCEq}
{
\delta \mathcal{ C} = \rho \cdot  \delta b + \hat \rho \cdot \mathcal{C} \, ,
}
where we defined 
\eqn{hatrhodef}
{
\hat \rho=-\rho \cdot \delta X \cdot \rho  \, ,
}
with $\delta X_{mn}$ an $(M \times 2N)$-dimensional matrix with components $\delta x^{(m)}_n$. Since $\delta X$ only depends on the leading $\lGB$ perturbation of the QNM wave functions, 
the matrix $\hat \rho$ does not depend on initial conditions. As announced, expression \eqref{deltaCEq} shows how coefficients $\delta \mathcal{ C} $ are directly determined via initial conditions. However, they do not depend solely on the initial conditions for the background anisotropy, i.e. coefficients $\mathcal{C}_n$, but also on the initial conditions for the perturbation of the metric, $\delta b_{\rm init}$. As we will show in a moment, the influence of a particular choice of $\delta b_{\rm init}$ on the pressure anisotropy is small.

\begin{figure*}[t]
\centering
\includegraphics[height=0.29\textwidth]{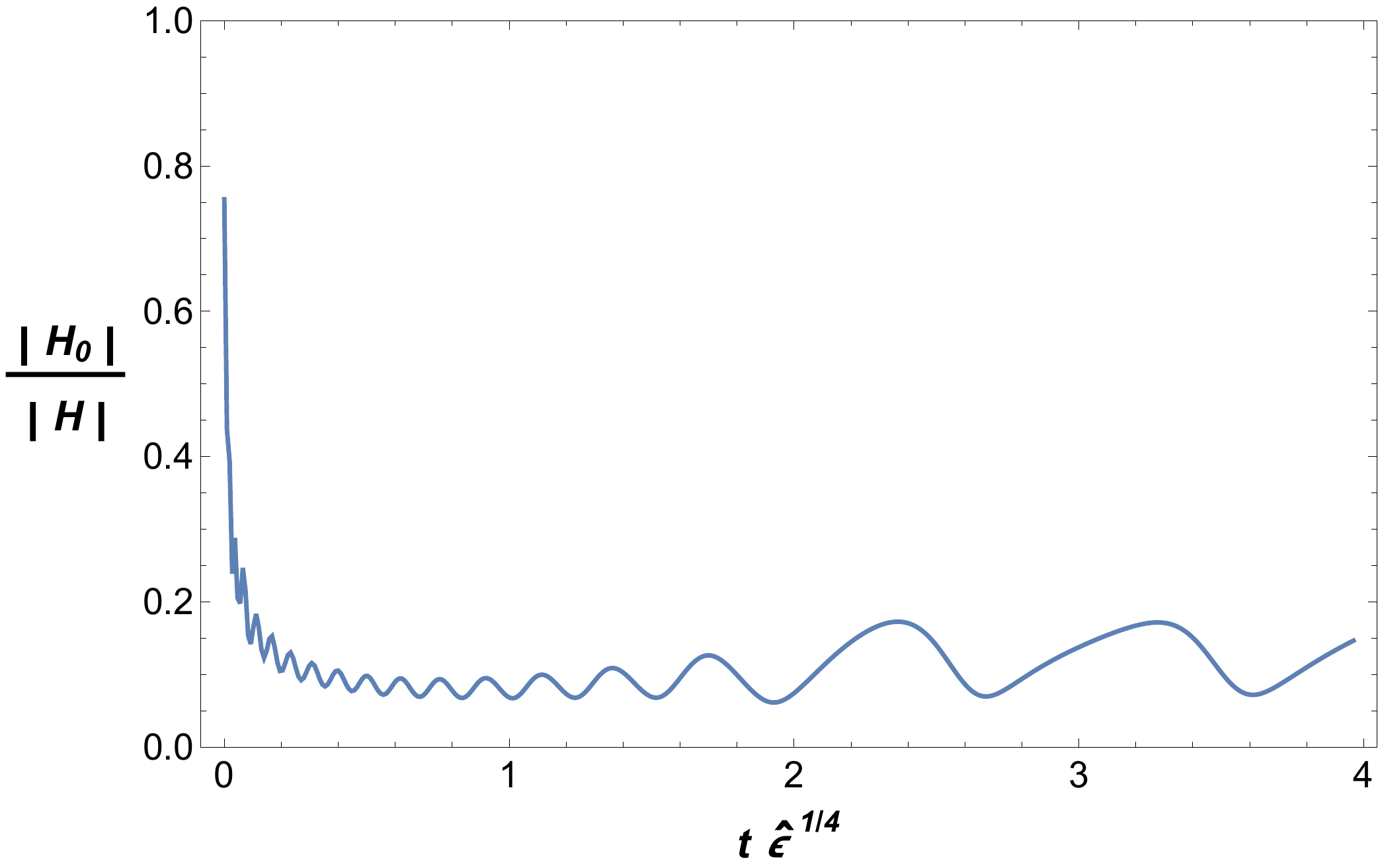}
\hskip 0.02\textwidth
\includegraphics[height=0.29\textwidth]{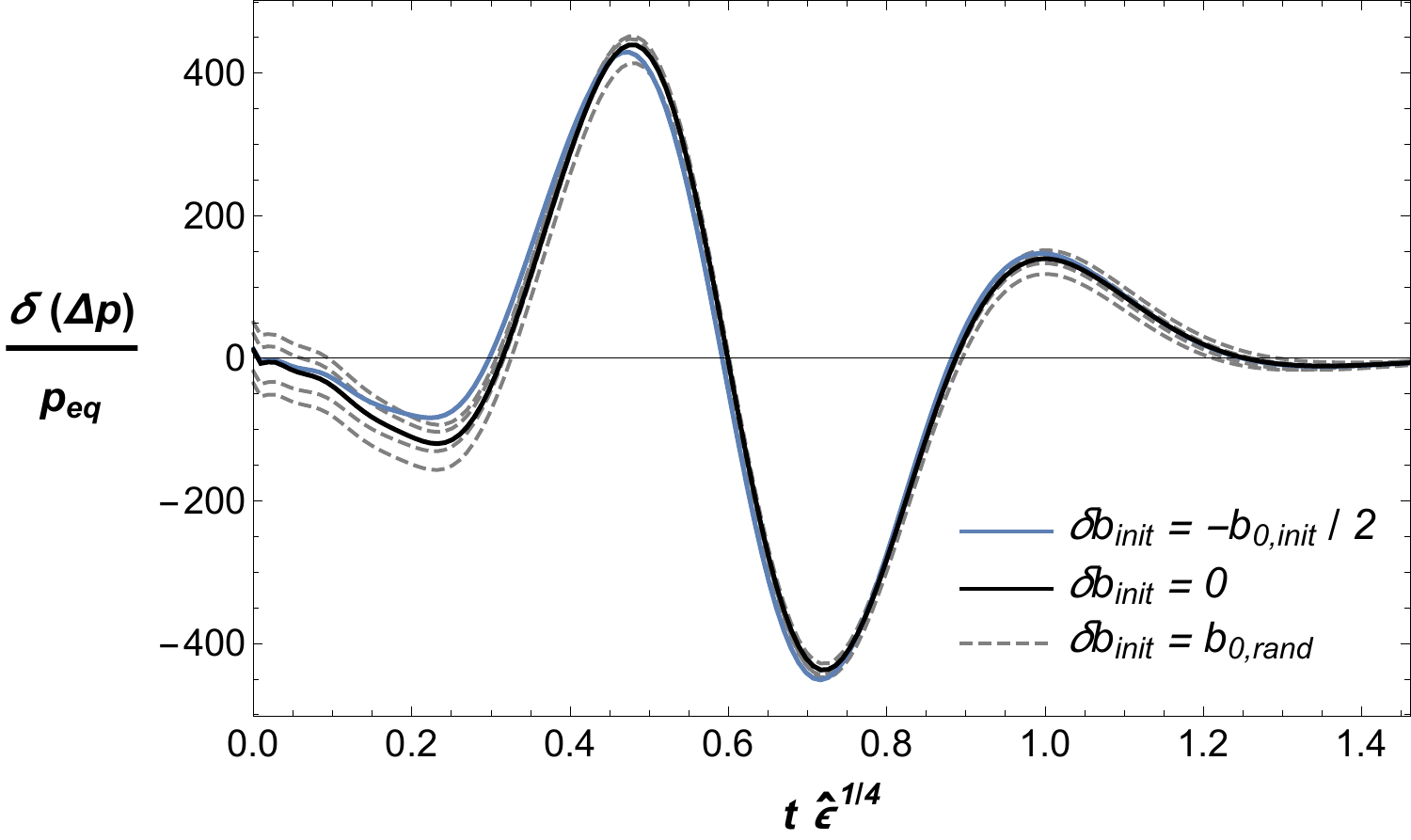}
\caption{Left: Ratio of the norms of the initial-condition independent functions $H_0$ and  $H$ from \eno{deltaDeltap3a}. For most of the time evolution, the norm of $H_0$ is much smaller that $H$. Right: comparison of the time evolution of the perturbed pressure anisotropy (as given by the QNM expansions) for several different initial conditions for the perturbations: $\delta b_{\rm init}=0$, $\delta b_{\rm init}=-b_{0,\rm init}/2$, and $\delta b_{\rm init}=b_{0, \rm rand}$, the latter being the rest of the initial conditions studied in Fig. \ref{p1}. } 
\label{plot:insensitivity}
\end{figure*}

With explicit expressions \eqref{SolRegression}, we may now directly relate the pressure anisotropy to the initial conditions for the metric. Before doing so, let us introduce some more notation to Eqs. \eqref{DeltaDeltaPExp} and \eqref{DeltaPExp}:
\eqn{deltaDeltap1} 
{\begin{aligned}
\Delta \hat p_0 &=-3 \, {\rm Re}\left[ \sum\limits_{n=1}^N e^{-i\omega_n^{(0)}t}
                                                                            D_n C_n^{(0)} 
                                                                 \right] \, ,\cr
\delta\left(\Delta \hat p\right) &=-3 \, {\rm Re}\left[ \sum\limits_{n=1}^N e^{-i\omega_n^{(0)}t}
                                                                           \left( K_n(t) C_n^{(0)} + D_n\delta C_n \right)
                                                                 \right]  \,,
\end{aligned}}
where the definitions of $K_n(t)$ and $D_n$ follow directly from Eqs. \eqref{DeltaDeltaPExp} and \eqref{DeltaPExp}. Note that these two sets of quantities are independent of initial conditions and are completely determined by the QNM of the late time solution, and that, since  $\Delta \hat p_0$ and $\delta\left(\Delta \hat p\right)$  only depend on the near boundary behaviour of the metric function, these variables do not depend on the holographic coordinate $u$. The final bit of bookkeeping consists of defining:
\eqn{BookKeeping2}
{\begin{aligned}
&\mathcal{T}_n = e^{- \Gamma^{(0)}_n t}\left(\cos \left( \Omega^{(0)}_n t \right) {\rm Re} \,K_n+ 
\sin \left( \Omega^{(0)}_n t \right) {\rm Im} \,K_n\right)\,,\cr
&\mathcal{T}_{N+n} = e^{- \Gamma^{(0)}_n t}\left(-\cos \left( \Omega^{(0)}_n t \right) {\rm Im} \,K_n+ 
\sin \left( \Omega^{(0)}_n t \right) {\rm Re} \,K_n\right)\,,\cr
&\delta \mathcal{T}_n = e^{- \Gamma^{(0)}_n t}\left(\cos \left( \Omega^{(0)}_n t \right) {\rm Re} \,D_n+ 
\sin \left( \Omega^{(0)}_n t \right) {\rm Im} \,D_n\right)\,,\cr
&\delta \mathcal{T}_{N+n} = e^{- \Gamma^{(0)}_n t}\left(-\cos \left( \Omega^{(0)}_n t \right) {\rm Im} \,D_n+ 
\sin \left( \Omega^{(0)}_n t \right) {\rm Re} \,D_n\right)\,,
\end{aligned}}
where $n=1,...,N$ and where we have explicitly separated the real and imaginary parts of the QNM frequencies,
$\omega_n^{(0)}=\Omega^{(0)}_n- i \Gamma_n^{(0)} $. Finally, we plug Eqs. \eqref{SolRegression} and \eqref{deltaCEq} into the QNM decompositions for the time evolution of the background pressure anisotropy and its perturbation, Eqs. \eqref{DeltaPExp} and \eqref{DeltaDeltaPExp}, and use the notation \eqref{BookKeeping2} to write them as:
\eqn{deltaDeltap3}
{
\begin{aligned}
\Delta \hat p_0 &=-3\left[H_{0}(t) \cdot b_0 \right] \, ,\cr
\delta\left(\Delta \hat p\right) &=-3\left[H(t) \cdot b_0 + H_0(t) \cdot \delta b \right]\,,
\end{aligned}
}
where 
\eqn{deltaDeltap3a}
{
\begin{aligned}
H_0(t)&\equiv \delta \mathcal{T}(t) \cdot \rho\,, \cr
H(t)&\equiv\mathcal{T} (t)\cdot \rho + \delta \mathcal{T} (t)\cdot \hat \rho\,,
\end{aligned}
}
are $M$-dimensional vectors with values at the grid points given by components $H_0^{(m)}(t)$ and $H^{(m)}(t)$\footnote{It is possible to take the continuous limit and write vectors $H_0(t)$ and $H(t)$ as smooth functions of $u$, i.e. $H_0(t,u)$ and $H(t,u)$, which would allow one to write \eno{deltaDeltap3} as an integral in $u$ over the initial condition.}, and $\mathcal{T}(t)$ and $\delta\mathcal{T}(t)$ are $2N$-dimensional vectors with components given in \eno{BookKeeping2}. We should point out that both $H_0(t)$ and $H(t)$ are collections of universal functions of time, independent of the initial conditions. In this way, \eno{deltaDeltap3} provides a direct link between initial conditions on the gravity side and the pressure anisotropy on the field theory side.

\begin{figure*}[t]
\centering
\includegraphics[height=0.29\textwidth]{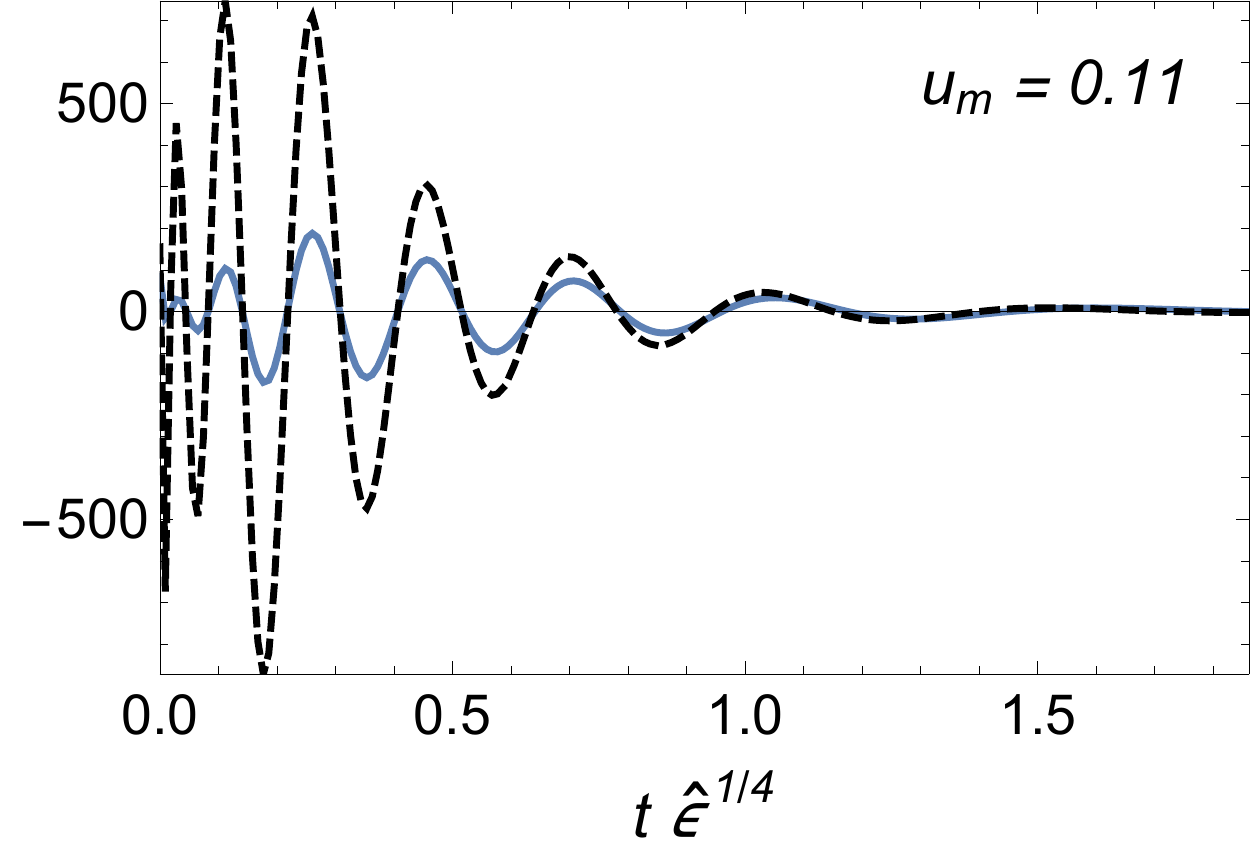}
\hskip 0.08\textwidth
\includegraphics[height=0.29\textwidth]{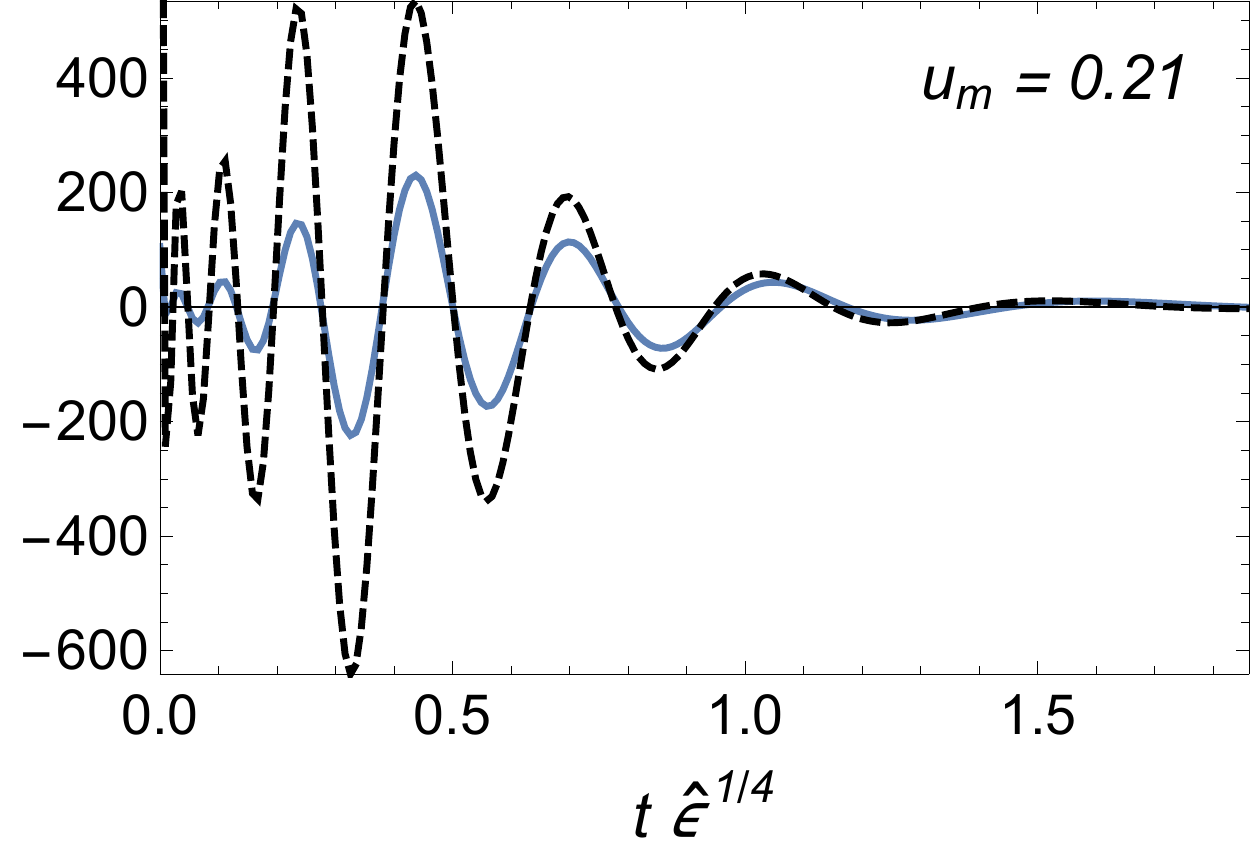}
\vskip 0.02\textwidth
\includegraphics[height=0.29\textwidth]{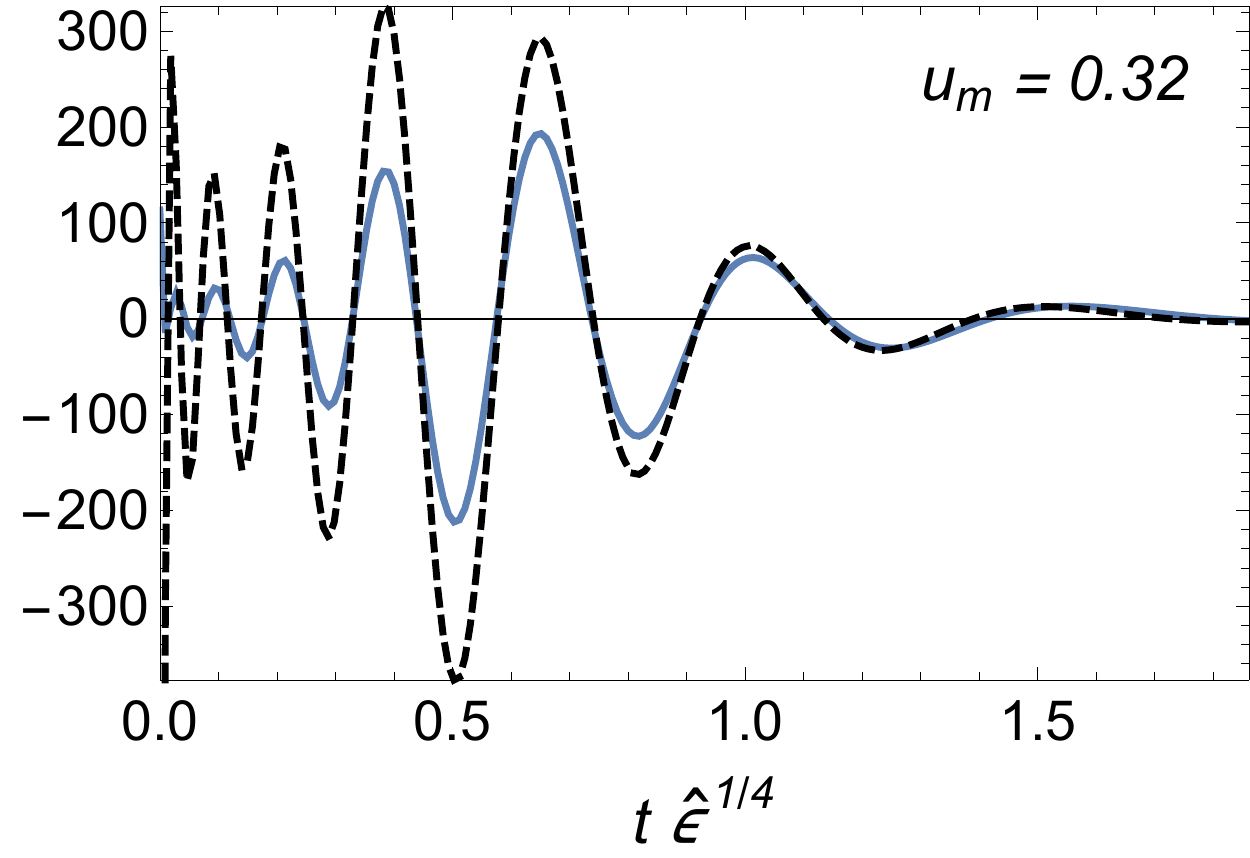}
\hskip 0.08\textwidth
\includegraphics[height=0.29\textwidth]{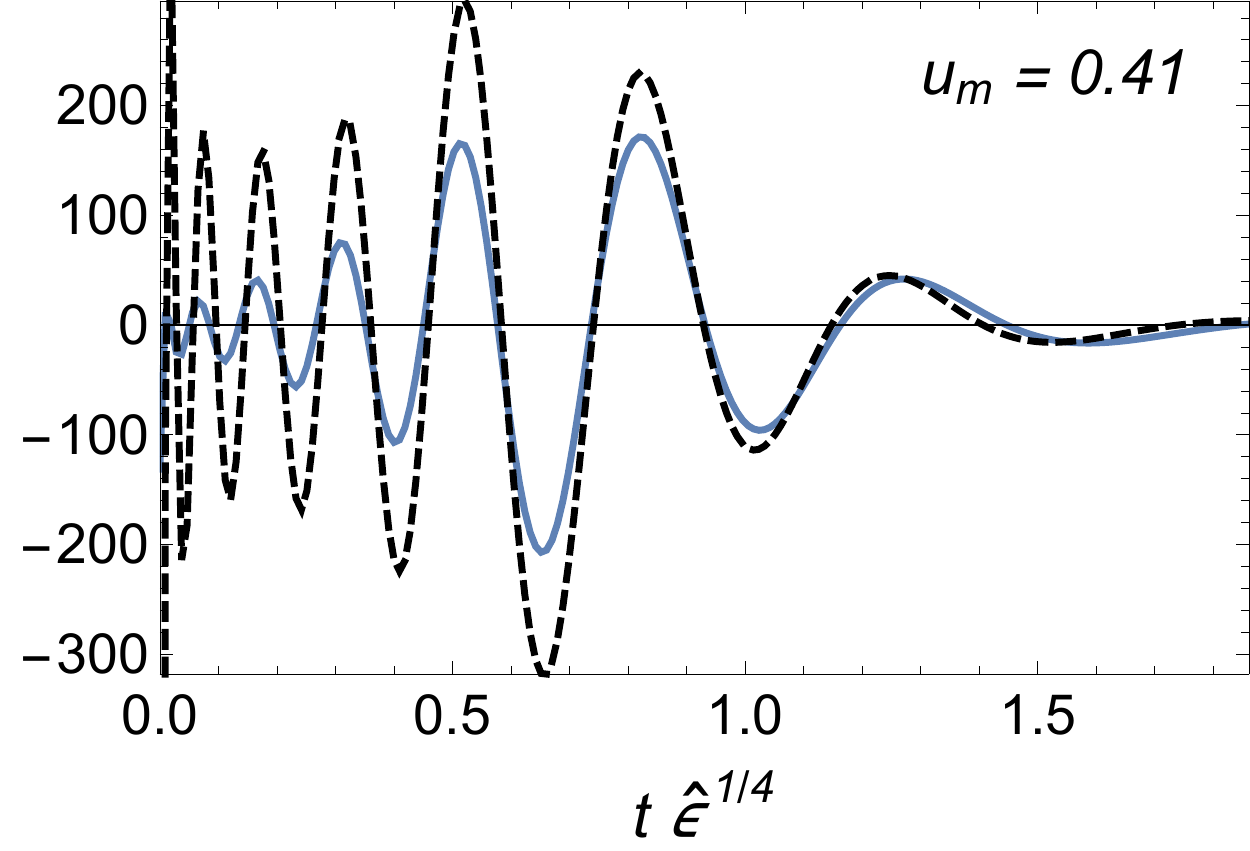}
\vskip 0.02\textwidth
\includegraphics[height=0.29\textwidth]{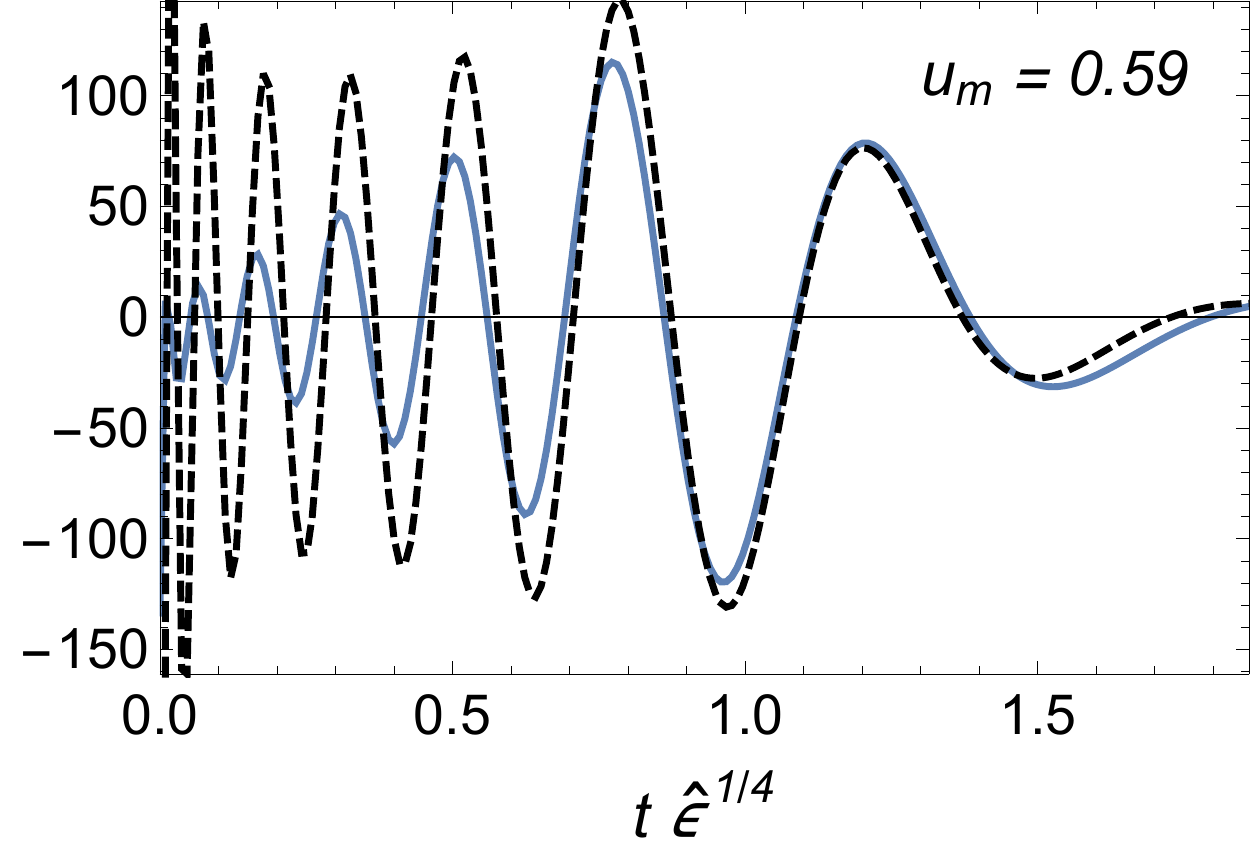}
\hskip 0.08\textwidth
\includegraphics[height=0.29\textwidth]{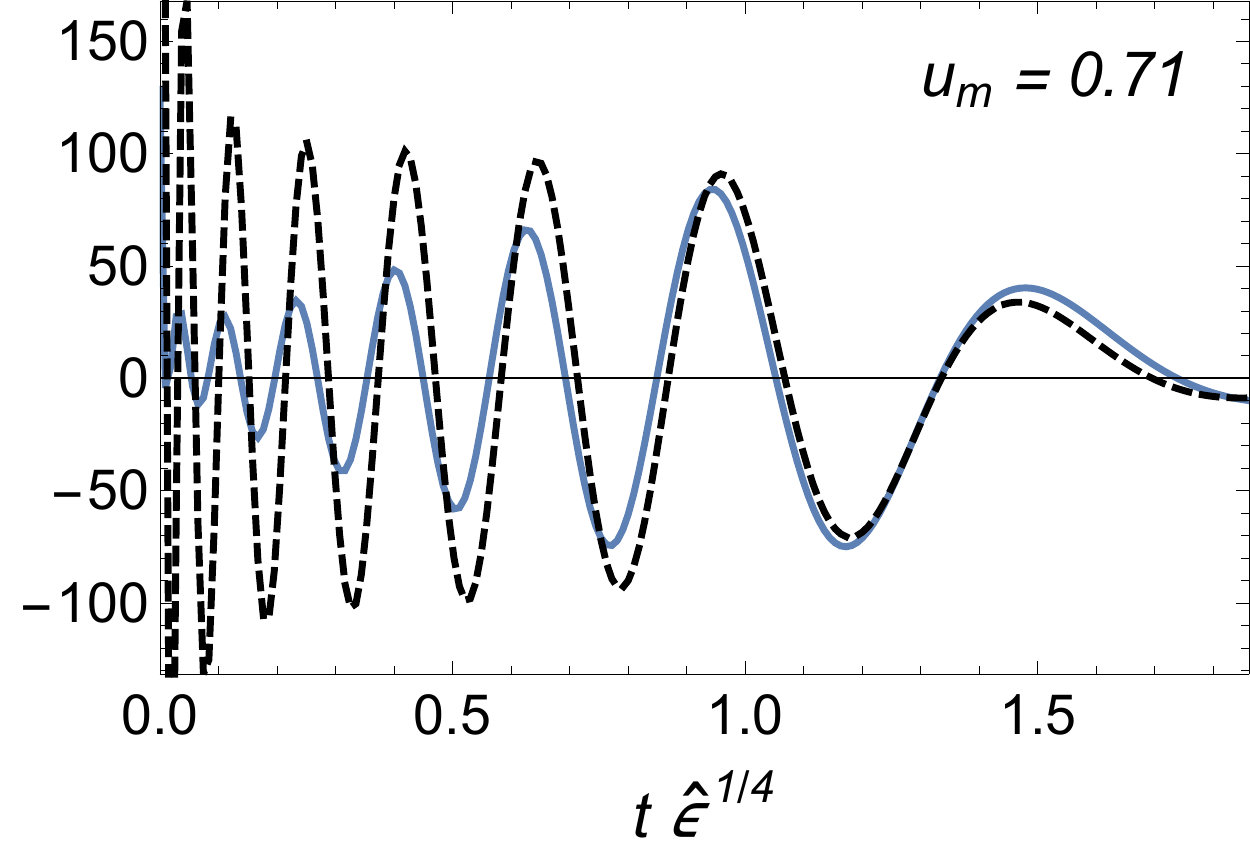}
\vskip 0.02\textwidth
\includegraphics[height=0.29\textwidth]{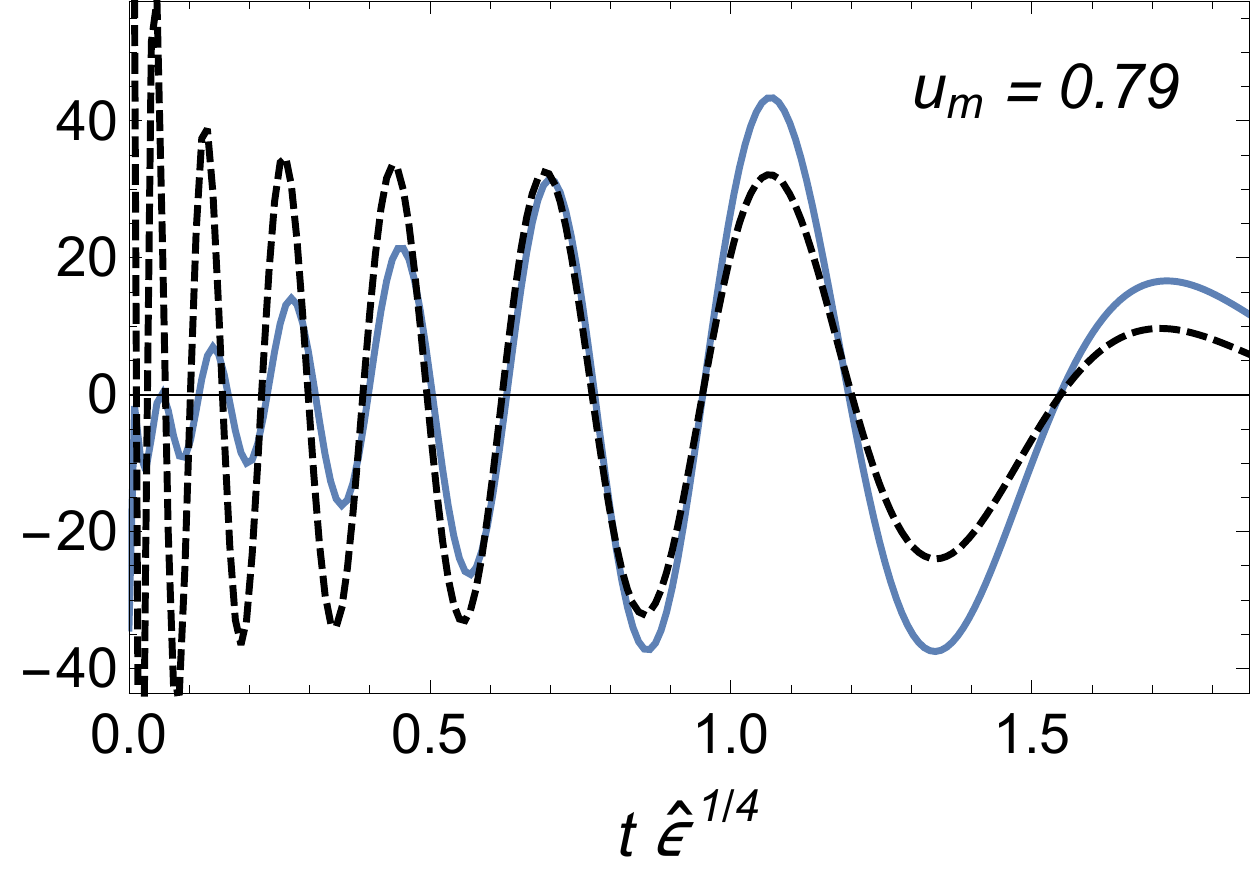}
\hskip 0.08\textwidth
\includegraphics[height=0.29\textwidth]{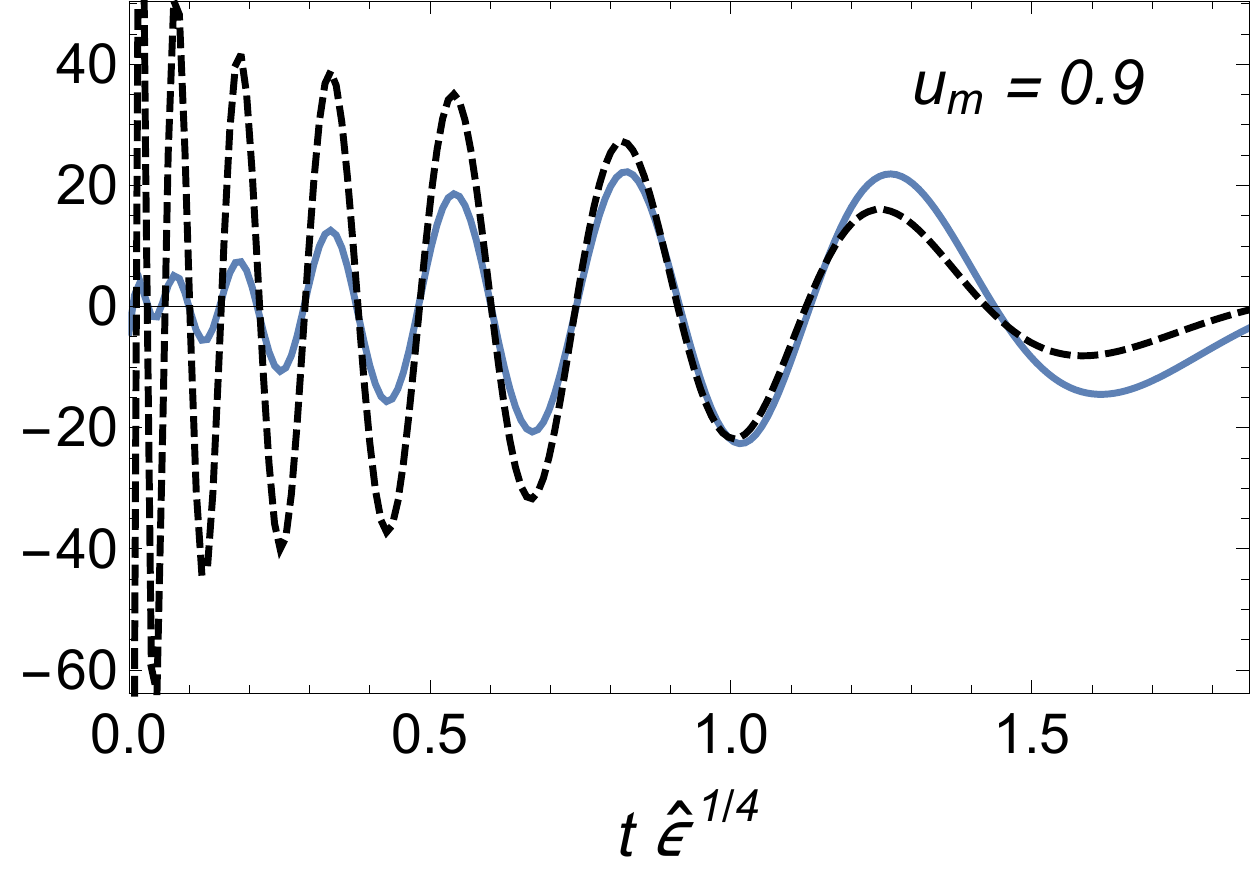}
\caption{Comparison of the initial condition-independent functions $H(t)$ (solid) and (the time derivative of) $H_0(t)$, $H'_0(t)$ (dashed), as function of time for different values of the holographic coordinate $u$. }
\label{plotH}
\end{figure*}

In \eno{deltaDeltap3}, the perturbed pressure anisotropy is expressed in terms of the initial conditions for the background metric field $b_0$ and the perturbation $\delta b$ via two sets of functions $H^{(m)}(t)$ and $H_0^{(m)}(t)$, which encode the time evolution of the anisotropy. Examining these two sets of functions numerically, we observe that the magnitude of the components of $H(t)$ is at least one order of magnitude larger than those of $H_0(t)$. To illustrate this fact, in the right panel of Fig.~\ref{plot:insensitivity} we show the ratio of norms of these two functions, defined simply as the square sum of the components over grid points, $\left| Z\right|^2 \equiv \sum  \left(Z^{(m)} \right)^2$.  As claimed, after a short transient time, the magnitude of $ H_0$ is one order of magnitude smaller than $H$ over the whole time range. This, in turn, means that as long as the magnitudes of initial conditions for $\delta b$ and $b_0$ are comparable, the influence of the exact form of the initial perturbation $\delta b_{\rm init}$ on the final perturbation of the pressure anisotropy $\delta(\Delta p)$ is small. 

To show this more explicitly, in the right panel of Fig. \ref{plot:insensitivity} we compare the perturbations of the pressure anisotropy (as given by the QNM expansion) evaluated for different choices of $\delta b_{\rm init}$. For the background initial condition $b_{0,{\rm init}}$ we chose the same one as in the first row of Fig. \ref{p1}, while the different choices of $\delta b_{\rm init}$ include our canonical choice  \eno{db IC}, and four sets of control initial conditions $\delta b_{\rm init}=0$ and $\delta b_{\rm init}= b_{0, \rm{rand}}$, where $b_{0, \rm{rand}}$ are the background initial conditions $b_{0,{\rm init}}$ from the rest of the panels in Fig. \ref{p1}. As we can see, the evolution of the perturbed pressure anisotropy becomes rather insensitive to the particular choice of $\delta b_{\rm init}$ already at early times. 
 
The observation above allows us to effectively neglect the second term in the expression for $\delta(\Delta p)$ in \eno{deltaDeltap3}, and describe the evolution of both the background pressure anisotropy $\Delta p_0$ and its perturbation $\delta(\Delta p)$ as linear combinations of values of only the background initial condition $b_{0,{\rm init}}$. Because of this convenient structure, we are now in position to demonstrate that the striking feature of our numerical analysis in Sec.~\ref{sec:time evolution GB} -- that the time evolution in the presence of Gauss-Bonnet terms can be understood as a time shift of the time evolution without it -- holds true for any initial condition $b_{0,{\rm init}}$, as long as the resulting time evolution of the pressure anistropy is approximately linear, so that the QNM expansion is valid.

As noted earlier, a necessary condition for the (nonconstant) time shift is that signs of $\Delta p_0$ and $\delta(\Delta p)$ are correlated, \eno{sign test for shift}. Due to the structure in \eno{deltaDeltap3} (without the second term in the expression for $\delta(\Delta p)$), this will hold true for any initial condition, as long as the signs of $H^{(m)}$ and $\partial_t(H_0^{(m)})$ are equal for all $t$ and $m$. As we can see in Fig. \ref{plotH}, this is indeed approximately true: the solid lines show the functions $H^{(m)}(t)$ for several fixed values of the holographic coordinate $u$ (fixed values of $m$), and the dashed lines show the time derivative of $H_0^{(m)}(t)$. For all values of $u$, after a short transient time these two initial-condition independent sets of functions become very similar, and satisfy \eno{sign test for shift} for most of the time range. This observation implies that for all those large (but not too large) pressure anisotropies which enjoy effectively linearised dynamics, the effect of the Gauss-Bonnet corrections may be understood as a time-shift of the isotropisation dynamics. As already stated in Sec.~\ref{sec:time evolution GB}, although the magnitude of this time-shift depends on the initial conditions, whether such shift is positive or negative is completely determined by the sign of $\lGB$.




\section{Exploring initial conditions}
\label{sec:discussion}

\begin{figure*}[t]
\centering
\includegraphics[height=0.29\textwidth]{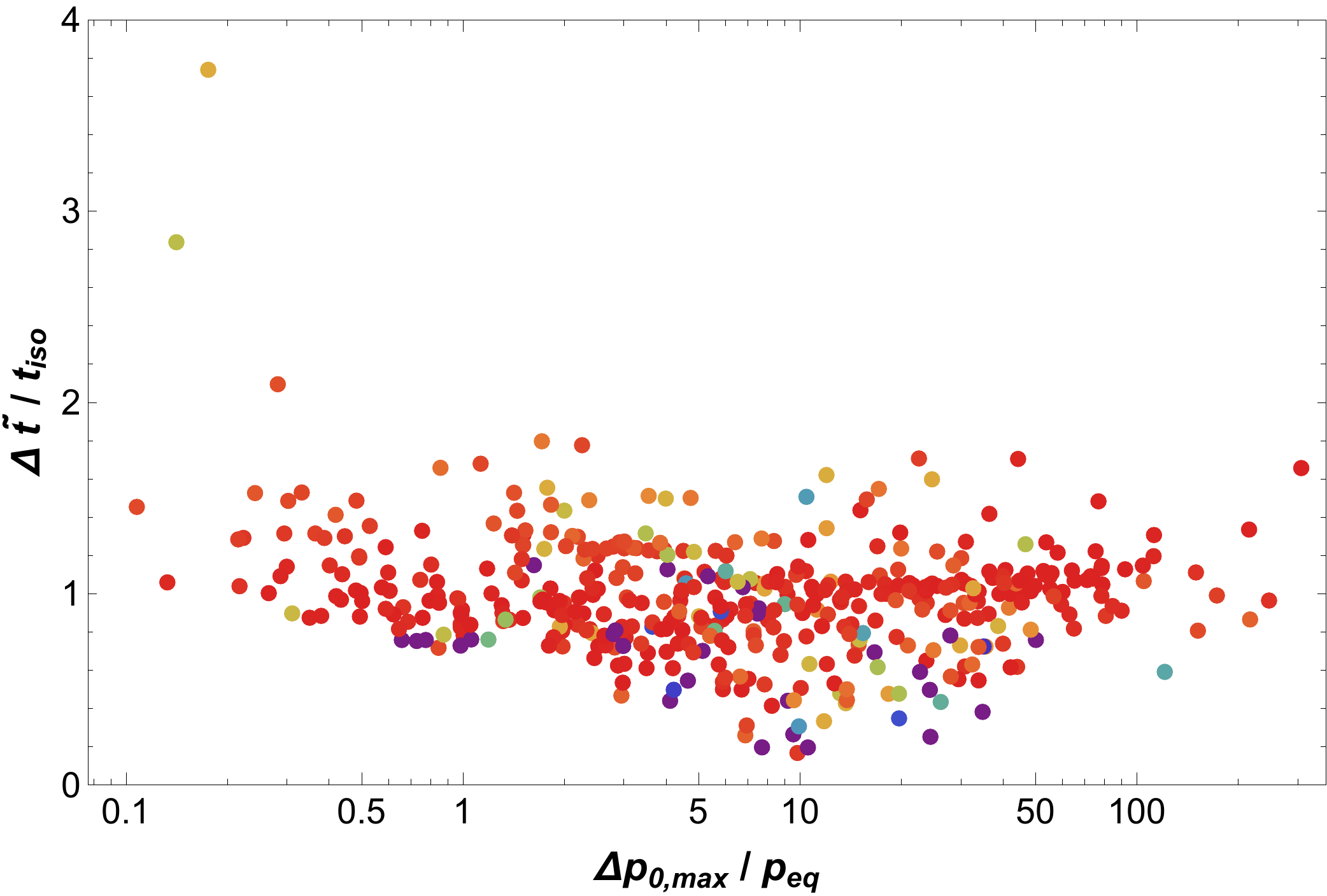}
\hskip 0.06\textwidth
\includegraphics[height=0.29\textwidth]{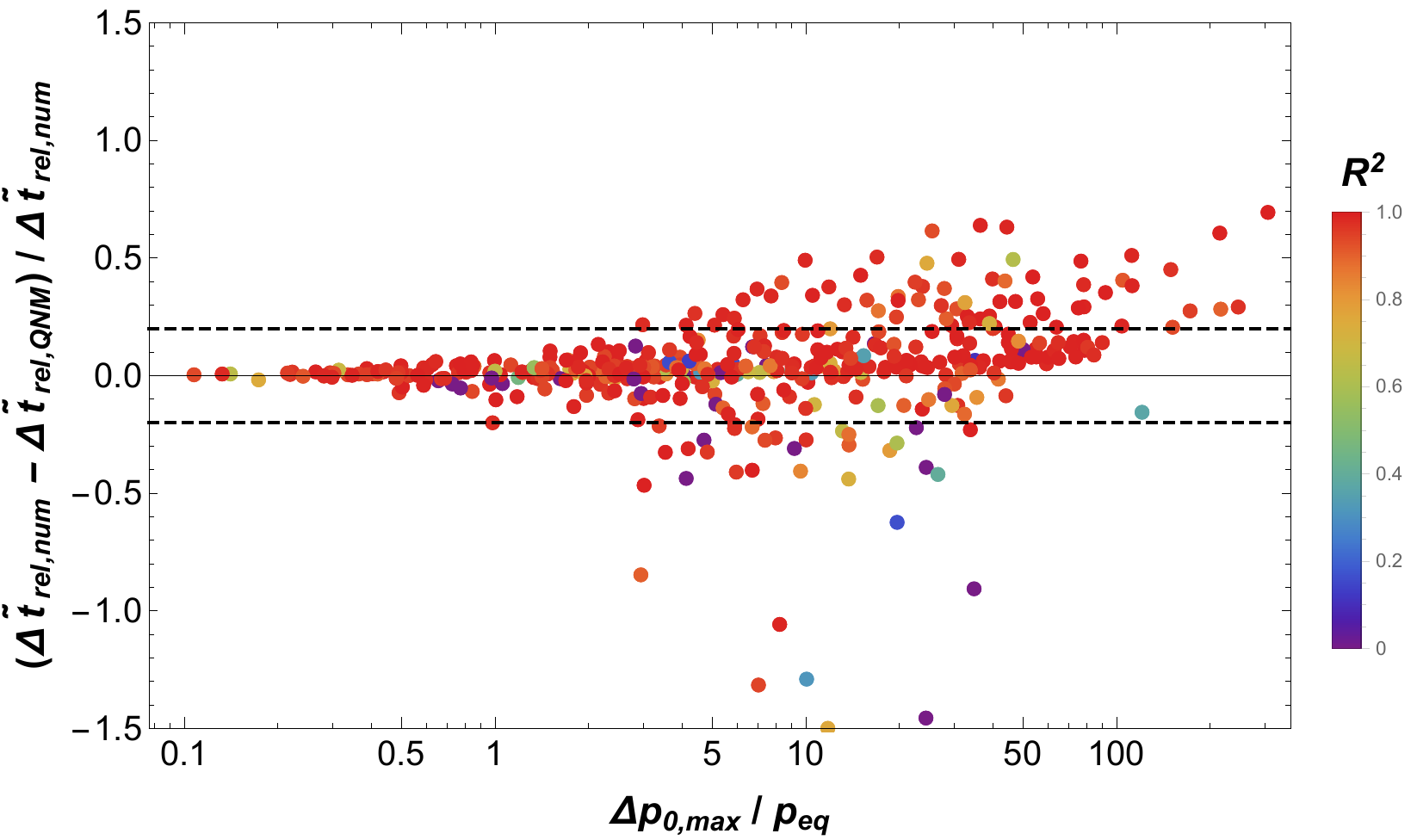}
\caption{Left: Best-fit values for time shifts $\Delta \tilde t$ normalized by the isotropisation times $t_{\rm iso}$ for the numerical backgrounds considered vs. the corresponding maximal values of the background pressure anisotropies. Points are color coded with $R^2$ values of the linear fit \eqref{constantshift} used to determine the best-fit values of $\Delta \tilde t$. Right: Normalised difference between the relative best-fit values for time shifts $\Delta \tilde t_{\rm rel}\equiv \Delta \tilde t /t_{\rm iso} $ from the numerical simulations and the values obtained from their corresponding QNM expansions vs. the maximal values of the background pressure anisotropies. Color coding is the same as in the left plot, and the dashed lines indicate values of $\pm 0.2$.} 
\label{plotShiftgen}
\end{figure*}

As we have seen, both from the  of numerical simulations considered so far and the analysis of the QNM expansions, the effect of a (negative) $\lGB$ correction leads to a delay in the isotropisation time. In this section, we will inspect how general this observation is by exploring the perturbed dynamics of 460 different random numerical configurations (generated using the procedure explained in Sec. \ref{preliminaries}), which we present in Fig. \ref{plotShiftgen}. 

For a large portion of those configurations, we have observed that  for  times $t>t_{\rm iso}$, with the isotropisation time $t_{\rm iso}$ defined as the time after which $\Delta p_0 /p_{\rm eq} < 0.1$, the effect of $\lGB$-corrections can be understood as an approximately constant shift of the dynamics. To see this, we have, following the discussion around \eno{sign test for shift}, fitted the numerical results for the perturbed anisotropy to the time derivative of the background anisotropy times a constant, $\alpha$:
\eqn{constantshift}
{\delta (\Delta p)(t) = \alpha \,\Delta p_0' (t)\quad {\rm for}\quad t>t_{\rm iso}\,,} 
and used linear regression to extract the best-fit value for the constant time shift $\Delta t = \lambda_{GB}\,\alpha\equiv\lambda_{GB}\Delta \tilde t $. The left plot in Fig. \ref{plotShiftgen} shows the ratio of $\Delta \tilde t / t_{\rm iso}$ as a function of the maximum background pressure anisotropy for all the numerical configurations we considered. The points are color coded with the value of the coefficient of determination $R^2$, a measure of how good the simple linear fit \eqref{constantshift} is. As we can observe, for a large portion of the simulations, \eno{constantshift} is a rather good fit: 83\% of the simulations have $R^2>0.8$, which signals a satisfactory fit. We should note that even the simulations with lower $R^2$, irrespective of the value of the maximum anisotropy, still display a time shift, albeit not a constant one. Another striking feature of Fig. \ref{plotShiftgen} is that, apart from a handful of low-$R^2$ outliers, the vast majority of points are spread over a relatively narrow time interval with $\left. \Delta \tilde t /t_{\rm iso}\right|_{\rm mean}=0.98$  and a standard deviation of $\sigma_{\Delta \tilde t / t_{\rm iso}}=0.32$.

Another key result of this work is the apparent effective linearisation of the far-from-equilibrium dynamics of the perturbation of the pressure anisotropy for large (but not arbitrarily large) background anisotropies, which admits a description in terms of a QNM expansion. 
 To quantify when the non-linearities become important
we have inspected the effect of describing the time evolution of the pressure anisotropy (both the background and the perturbed one) with QNM expansions on our observable of interest, the shift in the isotropisation time. 
In the right panel of Fig.~\ref{plotShiftgen} we show the difference between the relative time shift $\Delta \tilde t_{\rm num} / t_{\rm iso}$ obtained from the numerics and the one obtained from the corresponding QNM expansions, $\Delta \tilde t_{\rm QNM} / t_{\rm iso}$. 
This plot shows that for values of the maximum background pressure anisotropy $\Delta p_0/p_{\rm eq}<3$, the difference between the numerical result and the linearised approximation is less than 20\%. For larger values, however, the spread of this difference is larger and, even though there are some sets of initial conditions which can be nicely described by QNM, the quality of the description becomes worse for more general initial conditions.\footnote{We should note that the isotropisation time $t_{\rm iso}$ is a rather sensitive observable, and what may seem like large discrepancies between the numerics and the QNM results in the right plot in Fig.~\ref{plotShiftgen}, often are not. This was noted already in \cite{HoloLinJHEP}, and has to do with the fact that around times of the order of $t_{\rm iso}$, the pressure anisotropies become rather oscillatory and small differences between the numerics and the corresponding QNM approximation may sometimes lead to large differences between $t_{\rm iso}$.} Even though the values of maximum background pressure anisotropy for which the perturbed dynamics is well described by the QNM are generally lower than the values for which the background dynamics is well described, they are still rather high and in the regime in which we may not normally expect the linearised approximation to hold.

\section{Conclusions}
\label{sec:conclusions}

We have performed the first analysis of higher curvature effects on far-from-equilibrium anisotropic dynamics in holography.
%
Although it is yet unknown what is the precise gauge theory dual to Gauss-Bonnet gravity (if any), in this first exploratory work we have focused on this model since it provides a simple setup in which to address the effect of quadratic curvature corrections, which are the leading order corrections expected in a generic holographic construction. 
In the particular case of finite coupling corrections of $\N=4$, the leading order quadratic corrections identically vanish and the first non-vanishing correction is the quartic one. Nevertheless, as recently pointed out in \cite{AndreiGB}, when the higher derivative corrections are small, the features of quasinormal modes of Gauss-Bonnet and $\N=4$ SYM at large but finite coupling are qualitatively similar. Since QNMs control the far-from-equilibrium dynamics of large (although not arbitrarily large) anisotropies, we expect that many of our results apply to the case of finite coupling corrections to $\N=4$ SYM, at least qualitatively. 

One of the main results of this work is the determination of the effect of the Gauss-Bonnet terms on the isotropisation dynamics of far-from-equilibrium plasmas. In all the numerical simulations we have studied in Sec. \ref{sec:discussion}, with pressure anisotropies spanning over several orders of magnitude, the inclusion of (negative) $\lGB$ corrections leads to a $\lGB$-proportional shift of the time evolution of the full pressure anisotropy, for the most part of the evolution in which the anisotropy remains large. This shift then implies a delay in the isotropisation time (see Figs. \ref{p2} and \ref{plotShiftgen}).
In Subsec. \ref{TimeShift}, we have demonstrated that, as long as the description of the perturbed dynamics is effectively linear, this observation holds true for any choice of initial conditions (Fig.~\ref{plotH}). 

Our finding that, for all the numerical configurations considered, $\Delta \tilde t$ is always positive, and hence the sign of the physical time shift $\Delta t$ is alway opposite to the sign of $\lGB$ nicely resonates with the corrections to transport properties in this theory, since negative $\lGB$ implies larger $\eta/s$ ratio and vice versa. Quite intuitively, gauge theories with higher viscosities, which may be viewed as less strongly coupled, possess longer isotropisation times. It would be interesting to analyse other examples of large curvature corrections to verify these findings.

Another main result of this work is the apparent effective linearisation of the far-from-equilibrium dynamics at strong coupling for large, but not arbitrarily large anisotropies. Such effective linearisation is one of the most remarkable features of the relaxation of far-from-equilibrium anisotropic states in holography. This linearisation has been observed in configurations with strong anisotropies \cite{HoloLinPRL,HoloLinJHEP} with or without the magnetic field and chemical potential \cite{Fuini:2015hba}; in the dynamics of baryon charge in the collisions of shocks \cite{Casalderrey-Solana:2016xfq}; and in non-relativistic holography \cite{Gursoy:2016tgf}. Those studies indicate that, at least in the large coupling limit, the dynamics of those far-from-equilibrium processes are tremendously simplified. 

In this paper we have performed the first steps towards understanding whether this simplification survives once corrections to those extreme limits are considered. Within the context of Gauss-Bonnet gravity we have found  that linearised dynamics provide a good description of the full far-from-equilibrium processes even when the initial anisotropies are large ($\Delta p/p_{\rm eq} \sim \mathcal{O} (1)$). This regime is of particular phenomenological relevance, since the initial anisotropies of the far-from-equilibrium dynamics in heavy ion collisions at RHIC and the LHC are of that order. For those anisotropies, we have shown that the dynamics of the system can be predicted by a QNM analysis, much like in the unperturbed $\lGB=0$ case \cite{HoloLinPRL,HoloLinJHEP,Fuini:2015hba}. However, such simplification does not extend to arbitrarily large values of the initial anisotropy, as illustrated in Fig.~\ref{p1}.

From the gravity point of view, the onset of the non-linear behaviour is controlled by the right hand side of \eno{EELin}, which is sensitive to the higher derivatives of the background metric. As a consequence, when the pressure anisotropy is pushed to very large values, the corrections induced by the higher curvature terms exhibit clear non-linearities, even though the unperturbed solutions still behave surprisingly linear.
 On the gauge theory since, our results indicate that 
 the dynamics of gauge theories at finite but large coupling become more non-linear than in the infinitely strongly coupled limit. 

A clear limitation of our approach is the lack of control over the small $\lGB$ corrections to the initial out-of-equilibrium state. As we have discussed, in our setup we compare states that, independently of the value of $\lGB$, have the same initial anisotropy in units of the background energy density. However, this is not enough to fully specify the state. On the gravity side this ambiguity translates into the choice of initial $\delta b(u)$. One possible way to constrain this function is to demand the higher point correlators of certain operators are identical; however, to fully constrain the initial data, one would need an infinite set of those. Another possibility is to prepare the initial state as the result of a rapid quench of some anisotropic deformation of the theory independently of the value of $\lGB$. This procedure fixes the initial state uniquely. Nevertheless, our study of different functional forms of the initial conditions shows that our results remain independent of the details of the initial disturbances, provided that these are comparable to the initial anisotropy profile $b_0$. For anisotropies such that the system behaves effectively linearly, the independence on the initial disturbance can be expressed by the smallness of the initial condition independent function $H_0(t)$ as compared to $H(t)$, see Subsec. \ref{TimeShift}. For these reasons, we expect our main conclusions to remain unchanged for a generic far-from-equilibrium excitation of the system. 
   
The method we have employed in this paper is easily extendable to other sets of higher curvature corrections, as long as these corrections are small. Such small higher-curvature corrections will then lead to small deviations on top of the non-linear relaxation of the initial far-from-equilibrium state without corrections. In particular, the left hand side of \eno{EELin} is common to all gravities with small higher derivative corrections, since it only depends on the details of the time evolution of the far-from-equilibrium background we are perturbing. The difference between the various sets of corrections resides in the explicit form of the right hand side of \eno{EELin}, which may be understood as a source for the equation of motion for small perturbations on top of the background solution. This source term depends on different curvature tensors of the dynamical background. It would be interesting to generalise our study to other sets of corrections, such as the quartic curvature terms dual to finite coupling corrections of $\N=4$ SYM. Note, however, that the higher the order of curvature corrections is, the higher the derivatives of the background contained in the source term are. Therefore, to implement the quartic curvature terms, the dynamical background should be determined to a higher precision than what our current implementation allows. For this reason, we postpone this interesting analysis for future work. 

\vskip 0.2cm
\noindent {\bf Note added:} As this paper was being finalised, we became aware of a related work by S. Grozdanov and W. van der Schee in which they analyse the effect of the Gauss-Bonnet term on the far-from-equilibrium dynamics of the debris of two shock-wave collisions in holography. 

\acknowledgments

We thank Pau Figueras, Sa\v{s}o Grozdanov, Andrei Starinets and Wilke van der Schee for useful comments and discussions. TA and AF were supported by the European Research Council under the European Union's Seventh Framework Programme (ERC Grant agreement 307955). JCS is a University Research Fellow of the Royal Society. We are grateful to the organisers of the conference ``Numerical Relativity and Holography'' held in 
Santiago de Compostela in July 2016 for their hospitality during the completion of this work.


\bibliography{IsoBib}{}
\bibliographystyle{myJHEP}

\end{document}